\documentclass[%
 aip,
 amsmath,amssymb,
 reprint,%
]{revtex4-1}

\usepackage{graphicx}
\usepackage{dcolumn}
\usepackage{bm}

\usepackage[utf8]{inputenc}
\usepackage[T1]{fontenc}
\usepackage{mathptmx}
\usepackage{etoolbox}
\usepackage{natbib}
\makeatletter
\def\@email#1#2{%
 \endgroup
 \patchcmd{\titleblock@produce}
  {\frontmatter@RRAPformat}
  {\frontmatter@RRAPformat{\produce@RRAP{*#1\href{mailto:#2}{#2}}}\frontmatter@RRAPformat}
  {}{}
}%
\makeatother
\begin{document}

\preprint{AIP/123-QED}
\title[]{Extracting the Dispersion of Periodic Lossless LC Circuits Using White Noise}
\author{Daryoush Shiri}
\affiliation{Department of Microtechnology and Nanoscience, Chalmers University of Technology, Gothenburg, Sweden}
\author{Andreas Isacsson}
\email{shiri@chalmers.se}
\affiliation{Department of Physics, Chalmers University of Technology, Gothenburg, Sweden}

\date{\today}

\begin{abstract}
The spectral energy density (SED) method is used to obtain the phonon dispersion of materials in molecular dynamics codes, \textit{e.g.}, LAMMPS. We show how the electric analog of the SED method can be done using commercial circuit simulators to find the dispersion of periodic lossless LC circuits. The purpose of this article is (a) to demonstrate how SED proves useful, should the analytic methods of calculating dispersion of a circuit render difficult \textit{e.g.}, due to nonlinearity or having large number of elements in each unit-cell, and (b) to show how the concepts like Brillouin zone (BZ), dispersion (or band structure), zone folding, gap formation, and avoided crossing can be taught to students of electrical engineering by highlighting the analogies between phonons and periodic circuits. This analogy also suggests that thermal devices, \textit{e.g.}, heat rectifiers can be simulated and understood using commercial circuit simulators. 
\end{abstract}

\maketitle


\section{\label{sec:level1} Introduction}

Analogies between dispersion bands of vibrating atoms in a solid, energy band diagrams of electrons, and dispersion of voltage or current waves in a periodic discrete or distributed circuit were presented first by L. Brillouin \cite{Brilluoin}. In the 1950s, Gabriel Kron, an engineer from General Electrics, also proposed how the finite-difference form of partial differential equations can be simulated on periodic lattices of RLC circuits in one, two, or three dimensions \cite{Kron}. The solution of the time-independent Schrödinger equation with potential wells by an array of periodic LC or CL circuits was presented by W. Shockley in his book\cite{Shockley}. Other examples of using circuit analogy include emulating the Korteweg - De Vries (kdV) equation and soliton formation in Toda lattices in which nonlinear circuit elements \textit{e.g.}, varactors are used \cite{Toda,Giambo1984}. \\
Moreover, the physics of topologically protected states in condensed matters also sparked a new spate of research where people simulate different Hamiltonians and many-body interactions by periodic circuits with different dimensionalities and interconnections \cite{Price2022,Dong2021,Lee2018}. \\
\indent Commercial circuit simulators were not available at the time when Kron and others were trying to build the physics on circuits, despite the insight and pedagogical value the circuit analogy offers. Today, the availability of strong commercial circuit simulators which can treat circuits of all complexities in both time and frequency domains and in both linear and nonlinear regimes is an incentive to seek simpler ways of teaching concepts related to band structure in the condensed matter physics as well as investigating new physics and understandings using electric circuits. This approach will help both physics and engineering students as they are introduced to circuit analysis in their freshman college course, but band structure is a more advanced topic covered in solid state or device physics in later years. \\
\indent This article shows how the concept of spectral energy density (SED) is implemented for periodic circuits. The SED method is widely used in the calculation of phonon dispersion of solids, \textit{e.g.}, graphene, and CNT \cite{Thomas2010,Larkin2014}. The molecular dynamics codes solve equations of motion of atoms under different thermodynamical conditions called thermostats \cite{THOMPSON2022,Gulp2003}. The equilibrated velocity of atoms as a function of time is then used to find quantities of interest, \textit{e.g.}, the density of state of phonons and thermal conductivity of the solid using the autocorrelation function of all atomic velocities. Furthermore, in the SED method, these time-recorded velocities are used to obtain the dispersion plots of phonons, \textit{i.e.}, the energy of phonons ($\hbar\Omega$) versus their wave vectors (q). This is useful in investigating which type of phonons are dominant contributors in the heat-conducting mechanism. The broadening of each branch is a measure of how fast phonons lose their energy as a result of scattering. From this, the lifetime of phonons can be extracted. The slope of the dispersion plots is the group velocity of phonons. Furthermore, form these two quantities (lifetime and group velocity), the mean free path of phonons is obtained. The mean free path shows which geometrical feature of the nanostructure affects heat transport \cite{Zou2008,Balandin2007}. \\
\indent Here we demonstrate how applying the SED method to a periodic circuit helps in finding the dispersion plots, \textit{i.e.}, frequency of propagating voltage/current wave ($\omega$) versus the wave vector ($k$). Dispersion of a circuit, like that of phonons, has a wealth of information about the group velocity of propagating waves and stopbands within the circuit. This is useful in designing periodic nonlinear circuits for parametric amplification of microwave signals in quantum computing, sensing, and astronomy \cite{Aumentado2020}. It also helps to find topologically protected bands in the circuit that emulates a given many-body Hamiltonian.
Teaching these analogies to electrical engineering, material science, and physics students is useful in increasing the student's grasp of concepts like BZ, dispersion, zone-folding, and dynamical matrix. Also, in the case of periodic circuits of complex topology \textit{e.g.}, those with nonlinear inductors (using Josephson junctions) or nonlinear capacitors (using varactor diodes), the SED approach proves helpful in obtaining the dispersion without resorting to lengthy derivations first. \\
\indent This article is organized in the following order. The SED formula is derived for a simple 1D solid, and its circuit analogy is presented using voltages/currents instead of the velocity of atoms. Then, the method of simulating the circuits in WRSpice® \cite{WRSpice} to compute the voltages/currents is explained. Afterward, four circuits are presented as examples. The dispersion of the circuits are calculated based on Kirchhoff's voltage/current laws (KVL and KCL) and compared with those obtained from SED. The circuits are periodic LC, and periodic CL (which shows negative dispersion or left-handedness). The third one has a unit cell with two C’s and L’s to teach the concept of zone folding and bandgap (or stopband). The last one is the periodic LC coupled to a ring resonator in each unit cell to show the concept of avoided-crossing or degeneracy lifting (relevant to the degenerate time-independent perturbation theory). Avoided crossing and the creation of gaps are important concepts in designing RF/microwave filters \cite{Martin2003} and improving the performance of superconducting parametric amplifiers \cite{Aumentado2020,Fadavi2023}. These concepts also have parallel analogs in optics, \textit{e.g.}, Bragg grating filters in fiber optics, optical amplifiers, and meta-material design \cite{Gutierrez2013,Collin2001}. The SED method presented here is also applicable to 2D and 3D periodic circuits.

\section{\label{sec:level1} Method}
The working principle of SED is based on Parseval’s theorem which both physics and engineering students learn early in their college education within the subject of Fourier transform. The theorem simply states that the energy content of a signal is distributed in its harmonic content and can be found by adding the squared amplitudes of its harmonics, \textit{i.e.}, Fourier components. Derivation of the energy density formula for phonons is based on Parseval’s theorem \cite{Thomas2010,Larkin2014}, and here we only state the formula without the proof:
\begin{widetext}
\begin{equation}
\rho_{E}(\boldsymbol{k},\omega)=\frac{1}{4\pi\tau_{0}N_T}\sum_{\alpha}\sum_{b}^{B}m_{b}|\int_{0}^{\tau_0}\sum_{n_{x,y,z}}^{N_T}\dot u_{\alpha}(n_{x,y,z},b,t)e^{(i\boldsymbol{k}\cdot\boldsymbol{r}(n_{x,y,z})-i\omega t)}|^{2}\;. 
\label{eq:SED1}
\end{equation}
\end{widetext}
For the phonon case, the parameters in the above equation are as follows. $N_T$ is the total number of unit cells in the periodic solid, $\tau_{0}$ is the total integration time which approaches infinity. $m_{b}$ is the mass of each atom in the unit cell. The summation over $\alpha$ runs over three different directions in the solid \textit{i.e.}, $x,y,z$. The index $b$ is the index of each atom in the unit cell, and $B$ is the total number of atoms in each unit cell. The inner summation runs over the number of unit cells and $n_{x,y,z}$ is the index of each unit cell. The wave vector is denoted by $\boldsymbol{k}$ (sometimes as $\boldsymbol{q}$), and $\boldsymbol{r}$ is the position in the 3D space wherein the phonon propagates. The position of each atom, $\boldsymbol{r}$, depends on $n_{x,y,z}$. The velocity of atom number $b$ along direction $\alpha$ in the unit cell $n_{x,y,z}$ as a function of time ($t$) is found from the first derivative of oscillation amplitude $u$ which is $\dot u_{\alpha}$.\\  
For a one-dimensional solid or circuit, the first summation does not exist. Also, the summation over the number of atoms is removed as in a 1D circuit each $LC$ section simulates one atom in every unit cell. The number of unit cells is the number of $LC$ sections, and the index $n_{x,y,z}$ is a number spanning 1 to $N_T$ in one direction. The wave vector $\boldsymbol{k}$ and position $\boldsymbol{k}$ reduce to $k_{x}$ and $x$ in 1D case. Therefore, equation (\ref{eq:SED1}) is simplified as follows
\begin{widetext}
\begin{equation}
\rho_{E}(k_x,\omega)=\frac{m_b}{4\pi\tau_{0}N_T}|\int_{0}^{\tau_0}\sum_{n_{x}=1}^{N_T}\dot u_{n_{x}}(t)e^{i\omega t} e^{ik_x\cdot x} dt|^{2}=\beta|\sum_{n_x=1}^{N_T} e^{ik_x\cdot x}\int_{0}^{\tau_0}\dot u_{n_x}(t) e^{-i\omega t}dt|^{2}
\label{eq:SED2}
\end{equation}
\end{widetext}
The summation and integral are interchanged by virtue of linearity, and pre-factors are absorbed in $\beta$ which is assumed to be 1 without loss of dispersion information. As can be seen in equation (\ref{eq:SED2}), the SED in this simplified form is two consecutive Fourier transforms. The inner-most transform is applied to the velocity to return the $U(n_x,\omega)$ and the second transform is in the $n_x$ domain, \textit{i.e.}, a discrete Fourier transform in space which returns $\rho_{E}(k,\omega)$. The wave vector $k_x$ is given by
\begin{equation}
k_{x}\cdot x=\frac{2\pi n_x}{N_\text{T}} 
\label{eq:SED3}
\end{equation}
where $n_x=0,1,2,…,N_\text{T}$. This means $\rho_{E}(k,\omega)$ is a two-dimensional matrix containing the discrete frequency spectrum. In one direction, it stores the wave frequency  ($\omega$), and in the other, it stores the wave vector values ($k_x$) which span $0$ to $2\pi$. For the periodic circuit, we assume the unit cell length to be one, $a_x=1$ , but note that the answers depend trivially on the value of $a_x$. The remaining steps are (1) injecting the white noise into the circuit, and (2) running transient analysis for a long time and collecting the currents of each branch. The current and voltage are electrical analogs of mechanical velocity and force, respectively \cite{Bertuccio2022}. Note that in molecular dynamics, the velocity of atoms are due to thermal noise, and in a circuit, the velocity is equivalent to current at each node and the fluctuation is due to the injected white (Nyquist) noise. 

The periodic circuits under study are simulated in the WRSpice® simulator from Whitley Research \cite{WRSpice}, but in principle, it can be done in any Spice-based circuit simulator with time-domain (transient) analysis. The unit cell of each circuit is written in the netlist as a sub-circuit. In the main netlist, 100 copies of the sub-circuit are connected in series. The input node is connected to a time-domain white noise voltage. The source parameters are standard deviation and mean voltage. 
The mean value is chosen to be 0 as the definition of white noise mandates.\\
The simulation is performed for a sufficiently long time such that the possible back and forth reflections in the long discrete circuit due to impedance mismatch at the boundaries are settled. Total simulation time and the time step are typically $t_{\text{tot}} = 2000~\text{ns}$ and $dt=5~\text{ps}$, respectively. Thereafter, the node voltages (or the branch currents) are saved in a matrix of $N_\text{r}\times N_\text{c}$ size in which the $N_\text{r}=100$ is the number of rows (circuit nodes), and $N_\text{c}= \frac{t_{\text{tot}}}{dt} = 4\times 10^5$, is the number of columns (time steps). \ref{fig:SchemLCCL} shows the LC and CL periodic circuits with node voltages and unit cells marked. The current going from each node to the ground via the parallel element is called the branch current. 

The first Fourier transform is performed using the FFT function in MATLAB. This FFT corresponds to the inner-most Fourier transform in the SED formula in equation (\ref{eq:SED2}). The result of this part is a new matrix where the number of rows is still 100 (the number of nodes), but the number of columns is now the number of frequency points. This is determined in the first FFT step, and it should be large enough to include the highest frequency component of the voltage.
After the first transform, a new FFT is performed in the vertical direction, \textit{i.e.}, in the spatial (node) domain. This FFT corresponds to the outer-most Fourier transform, \textit{i.e.}, the one over space (atom indices or node indices) in the SED formula [equation (\ref{eq:SED2})]. The matrix obtained in this step is visualized by surf or image functions in MATLAB, and the top view or the 2D rendering shows the trace of dispersion branches of the circuit. Each bright point on the branch corresponds to a peak. The sharpness of the peak depends on the number of points (pixels) along the $k$ axis, meaning that increasing the number of unit cells to 500 or 1000 can make the branches very fine and sharp at the expense of time and memory.\\ 
In the next section, we show four different periodic circuits for which the above method is used to obtain their dispersion. The results are compared with the dispersions found by analytic method based on Kirchhoff’s current and voltage laws and the eigenvalue problem, which is analogous to the dynamical matrix method for phonons \cite{Dove2003}. It is shown that the noise-based method presented here gives information about the dispersion, $\omega(k)$, and there is no need to solve and diagonalize the dynamical matrix.   

\section{\label{sec:level1} Examples}
\subsection{\label{sec:level2}One-dimensional periodic LC circuit}
\begin{figure}
\includegraphics[width=3.6in]{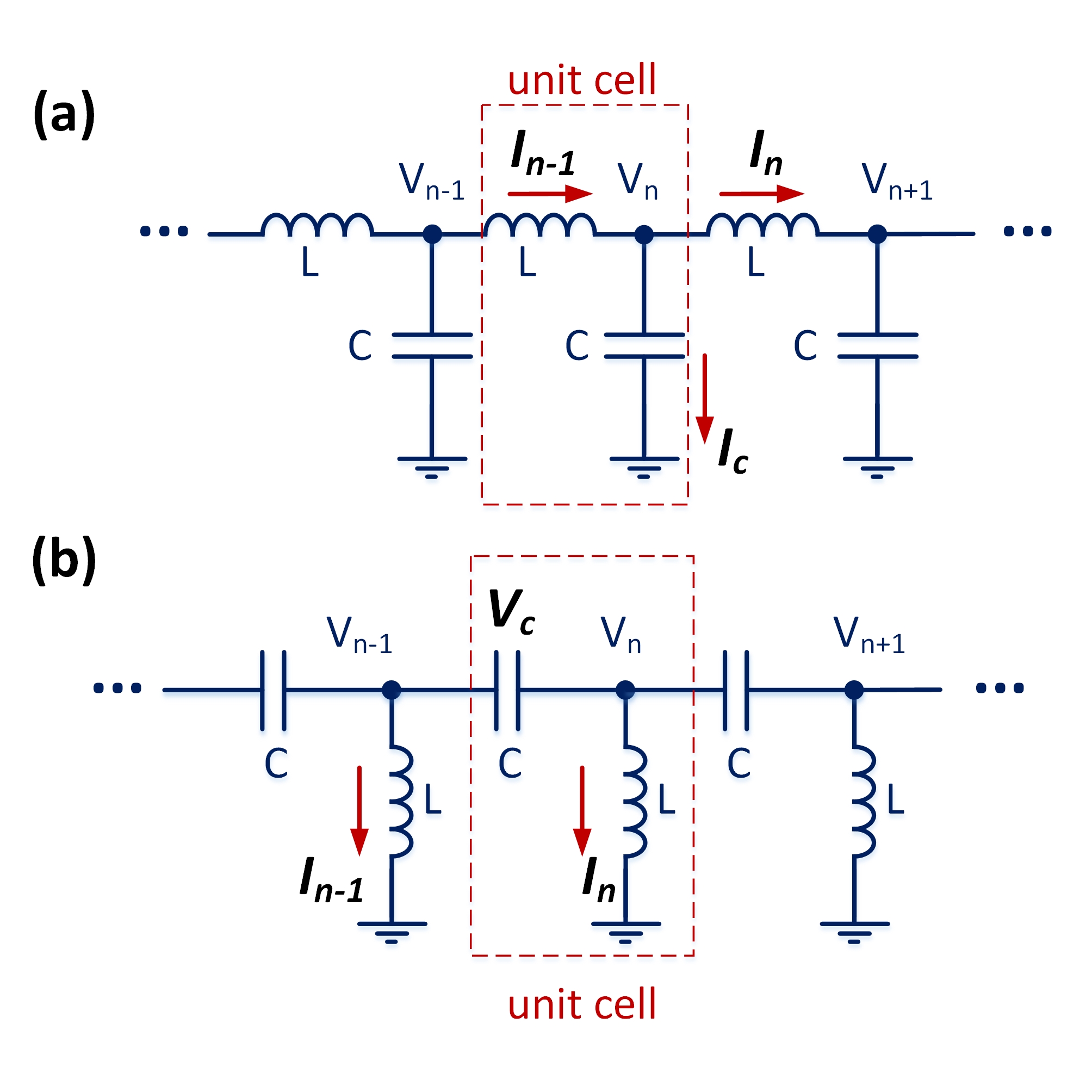}
\caption{The one-dimensional periodic circuits and their unit cells which is composed of (a) LC and (b) CL section. The circuit in (a) is a low-pass filter, and the one in (b) is a high-pass filter.}
\label{fig:SchemLCCL} 
\end{figure}

The one-dimensional infinite and periodic array of the equal inductors (L) and capacitors (C) is shown in Figure \ref{fig:SchemLCCL}(a). This circuit also serves as a discretized model of a continuous transmission line working at RF or microwave frequencies if we know the inductance per unit length ($l^{\prime}$) and capacitance per unit length ($c^{\prime}$) of the transmission line. We are interested in knowing which modes (waves of voltage or current) can propagate from $-\infty$ to $+\infty$ in this circuit with given inductance and capacitance determined by the geometry of the transmission line.
Akin to the method of treating periodic solids in solid-state physics \cite{Brilluoin}, we define a unit cell, a cell of the minimum size that can be repeated by many periodic shifts to build the whole circuit. In this case, the unit cell is composed of a parallel capacitor $C$ and a series inductor $L$, as shown in Figure \ref{fig:SchemLCCL}(a). The voltage of each node is $V_n$ and the current entering the $n$’th unit cell is $I_n$. Kirchoff’s current law (KCL) relates the currents in unit cell $n$ and its neighboring unit cell, $n-1$ as follows
\begin{equation}
I_{c}=I_{n-1} - I_{n} \rightarrow C\frac{dV_n}{dt}=I_{n-1} - I_{n} 
\label{eq:LC1}
\end{equation}
By taking time derivatives from both sides of equation (\ref{eq:LC1}) and noting that the relation between the magnetic flux inside the inductor and its current is $\Phi = \text{L}\cdot\text{I}$ and recalling Lenz’ law which is $V=-\frac{d\Phi}{dt}$, equation (\ref{eq:LC1}) is then converted to a second-order differential equation. This relates the voltages of node ($n$) and its next neighbors.
\begin{equation}
C\frac{d^2V_n}{dt^2}=\frac{1}{L}(V_{n-1} + V_{n+1} -2V_{n}),
\label{eq:LC2}
\end{equation}
To solve equation (\ref{eq:LC2}) a periodic trial solution is used by assuming that the solution, voltage at node $n$, is a propagating wave with frequency, $\omega$, and $k_xx = kna$ (where $a$ is the length of the unit cell) and complex amplitude of $V_0$. After replacing the trial solution $V_n=V_0 e^{i(\omega t-kna)}$ in equation (\ref{eq:LC2}), we get
\begin{equation}
-C\omega^2V_0 e^{i(\omega t -kna)}=\frac{1}{L}V_0 e^{i(\omega t - kna)}(e^{-ika}+e^{ika}-2)
\label{eq:LC3}
\end{equation}
After solving for the frequency, we have
\begin{equation}
\omega^2=\frac{4}{LC}\sin^2(\frac{ka}{2}) 
\label{eq:LC4}
\end{equation}

\begin{figure}
\includegraphics[width=3.5in]{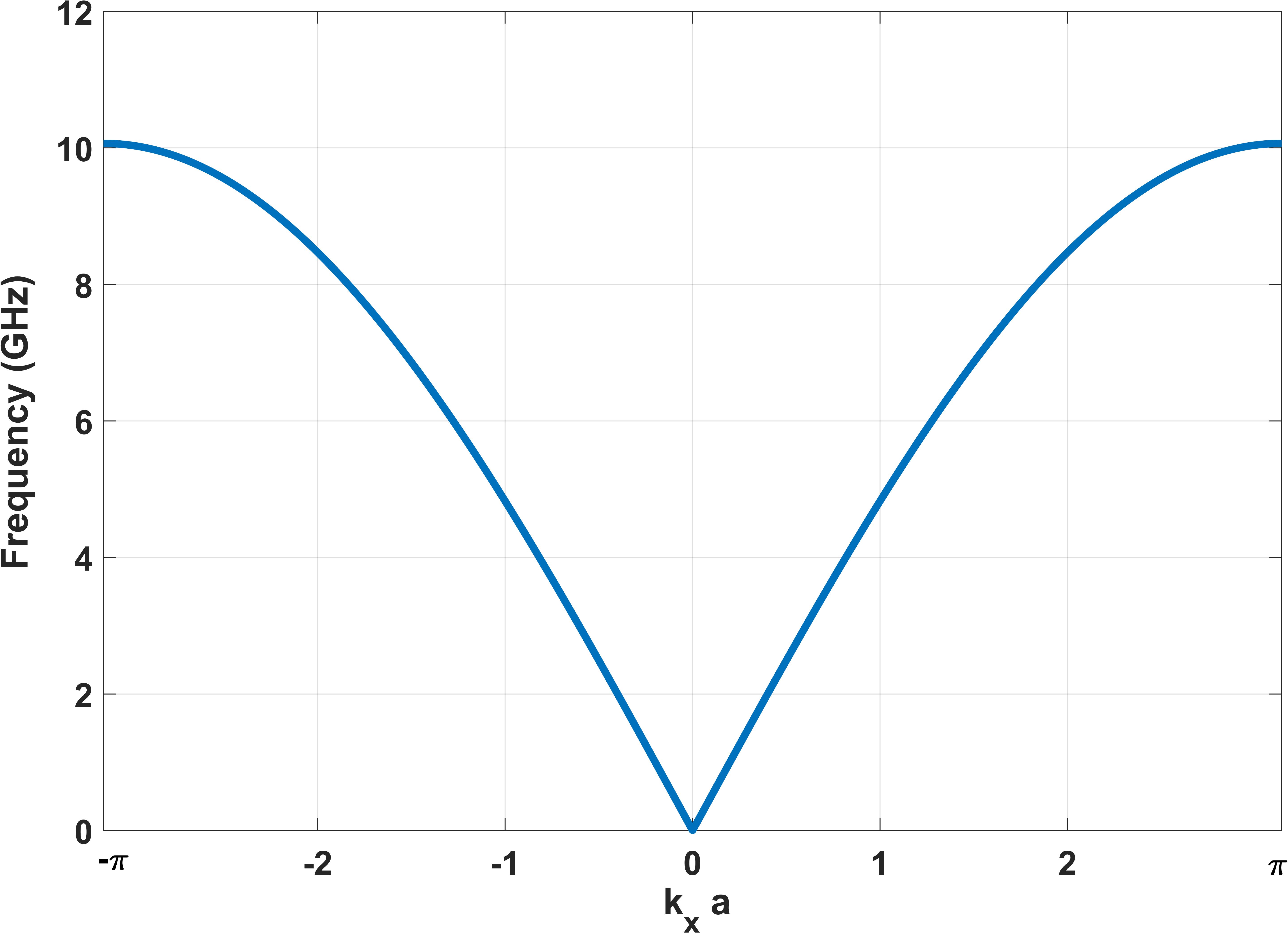}
\caption{ Dispersion of the 1D LC circuit shown in the first BZ ($- \pi \leq k_xa \leq \pi$) for $L = 1~\text{nH}$ and $C = 1~\text{pF}$.}
\label{fig:dispLC}
\end{figure}

\begin{figure*}
\centering
{\includegraphics[width=3in]{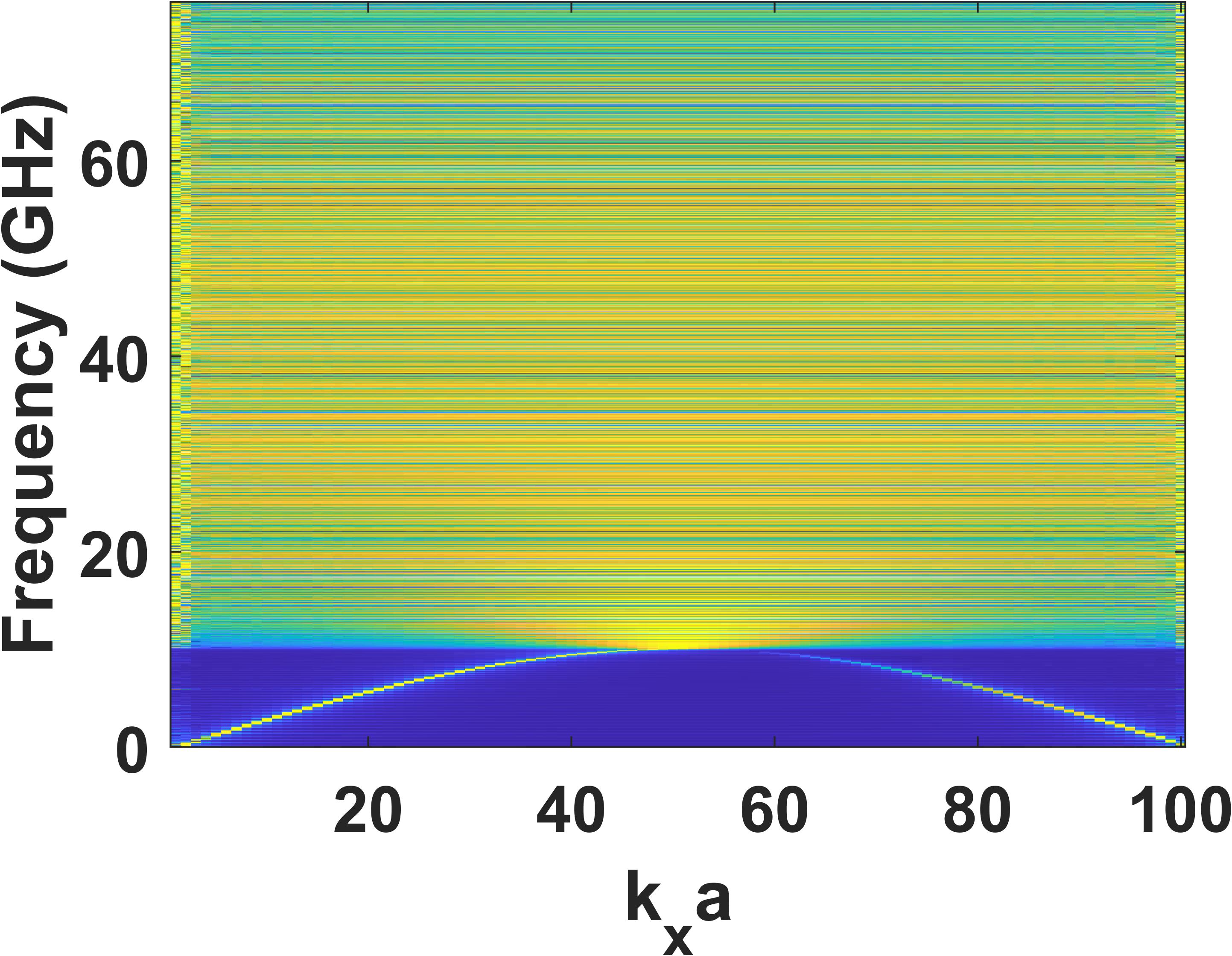}%
\label{fig(a)}}
\hfil
{\includegraphics[width=3in]{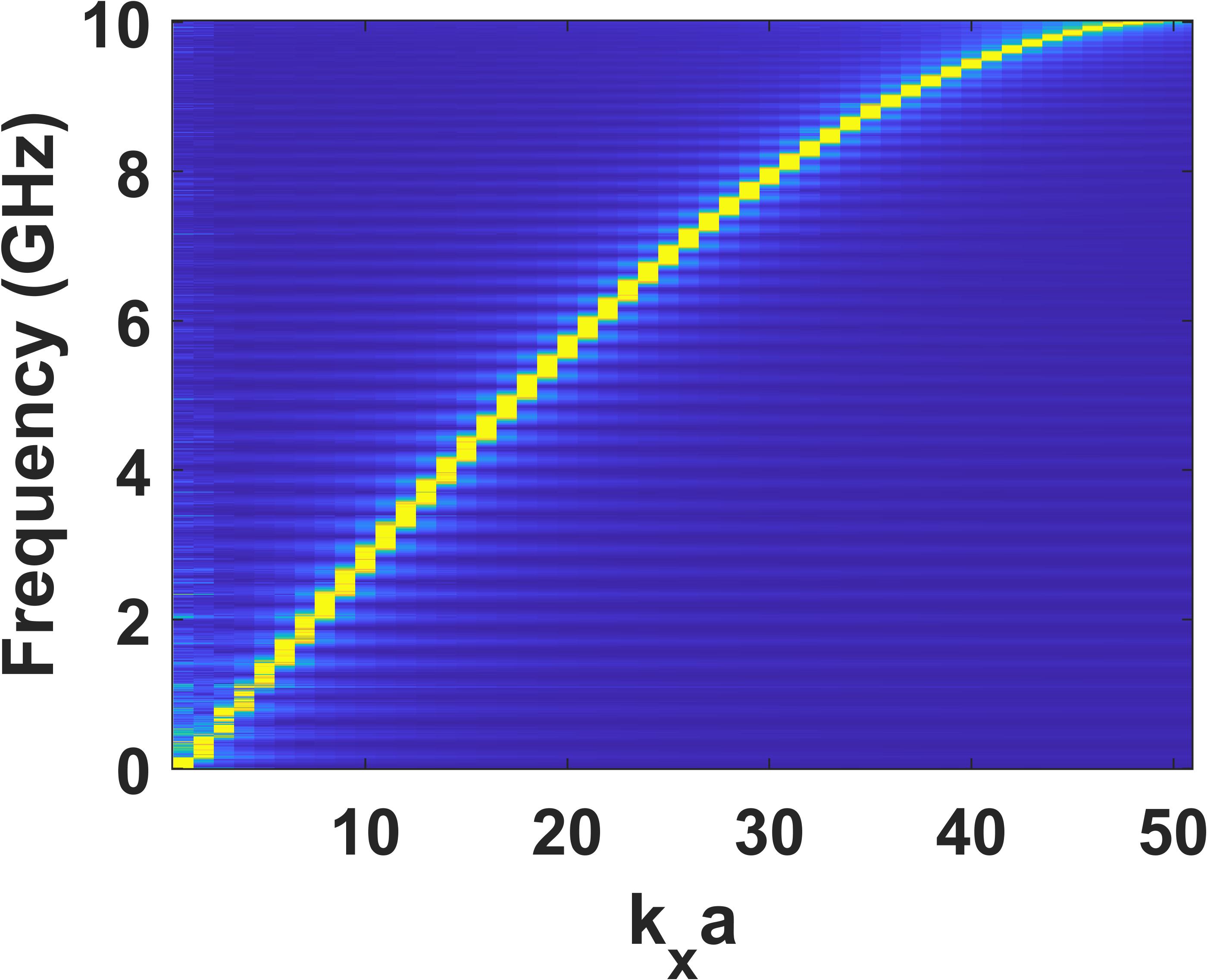}%
\label{fig(b)}}
\caption{Dispersion of the 1D LC circuit obtained by the SED method. (a) The 2D view of SED according to equation (\ref{eq:SED2}). Note that the data above the cut-off frequency of $10~\text{GHz}$ is noise. Below $10~\text{GHz}$, two replicas of branches are visible. (b) Half of the BZ with one branch. The number of points along $k_x$ is determined by the number of unit cells. Obtaining higher resolution requires a larger number of unit cells. Compare this with the right half of Figure \ref{fig:dispLC}.}
\label{fig:dispLCSED} 
\end{figure*}

Equation (\ref{eq:LC4}) has two solutions but the the positive one is meaningful. Plotting the frequency $\omega/2\pi$ versus the wave vector ($k$) or the normalized value ($k_x\cdot a$) yields the dispersion shown in Figure \ref{fig:dispLC}. Due to the periodicity of the dispersion, all information that we require to understand the dynamics of the circuit is contained in the interval of $k_xa = 2\pi$, \textit{i.e.}, $- \pi \leq k_xa \leq \pi$. This interval is the 1st Brillouin Zone (BZ). Everything outside the BZ is just a periodic replica of this information. 
The group velocity in the circuit is defined by $v_g=\frac{\partial \omega}{\partial k}$. This quantity shows the velocity by which a pulse (a wave packet) propagates in the circuit or the discrete model of the transmission line. In that case, each mode propagates with a different velocity if $\omega(k)$ is not linear. Then at the other end of the line, the pulse spreads or disperses and this is the reason behind the name, \textit{dispersion}. 
Note that at $k_xa = \pm \pi$, the dispersion is flat, \textit{i.e.}, the group velocity is zero. This suggests that the waves of this specific wavelength are standing waves. This resembles the Fermi level or Fermi surfaces in the solid's electronic band structure, which should always be perpendicular to the 1st BZ boundaries \cite{Brilluoin}. Also, the maximum frequency in the dispersion is about $10~\text{GHz}$, meaning that no mode with a higher frequency than this, can propagate. This frequency ($10~\text{GHz}$) is called the cut-off or bandwidth of the circuit, and it is obtained from equation (\ref{eq:LC4}) by putting $k_xa = \pm \pi$. For capacitor and inductor values of $1~\text{pF}$ and $1~\text{nH}$, respectively, we obtain,
\begin{equation}
\omega_{\text{cutoff}}=\frac{2}{\sqrt{LC}} \rightarrow f_{\text{cutoff}}=\frac{1}{\pi\sqrt{LC}}= 10~\text{GHz} 
\label{eq:LC5}
\end{equation}
Figure \ref{fig:dispLCSED} shows the dispersion obtained by the SED method for the 1D LC circuit. The circuit simulated in this method is composed of 100 unit cells [Figure \ref{fig:SchemLCCL}(a)]. Fourier analysis of noise currents according to equation (\ref{eq:SED2}) resulted in Figure \ref{fig:dispLCSED}(a). As the cut-off frequency was designed to be $10~\text{GHz}$, it is evident that there is no band (propagating mode) above $10~\text{GHz}$, and the results are noisy. If the circuit is sufficiently long, the forbidden modes have enough time to be attenuated by the circuit and the region above the cut-off frequency will be more noise-free. The first half of the 1D BZ (50 $k_x$ points) in Figure \ref{fig:dispLCSED}(b) has all information in one branch, which is the same as the right half of Figure \ref{fig:dispLC}.

Note that the second branch in Figure \ref{fig:dispLCSED}(a) is the mirror image of the first one, and it can be easily shifted to the negative $k_x$ values to make it the exact replica of Figure \ref{fig:dispLC}. However, one branch has all the necessary information about the circuit. 

\subsection{\label{sec:level2}One-dimensional periodic CL circuit}

\begin{figure}
\includegraphics[width=3.5in]{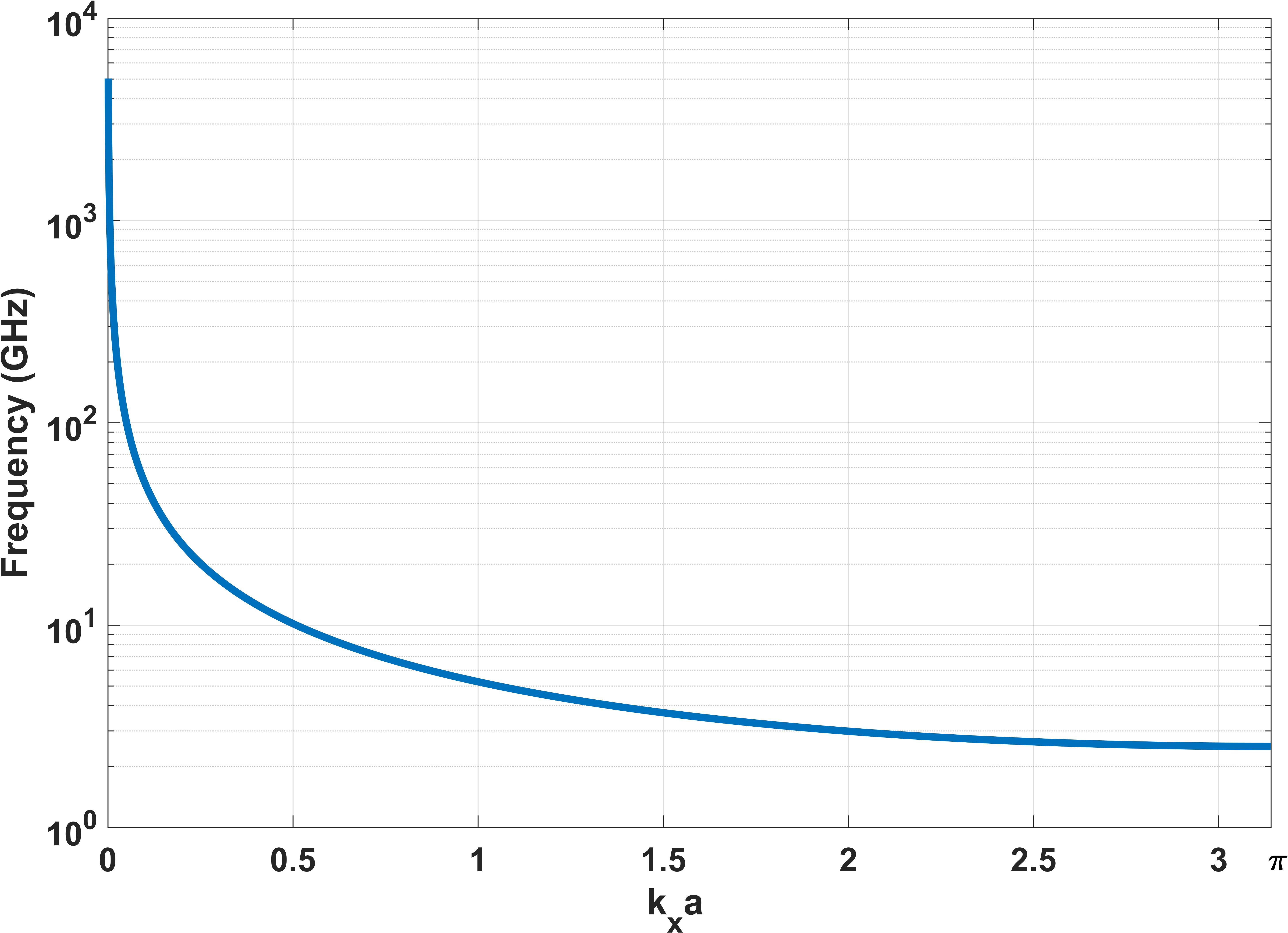}
\caption{The dispersion of 1D periodic $CL$ (left-handed) circuit for $L = 1~\text{nH}$ and $C = 1~\text{pF}$. The low cut-off frequency is $2.157~\text{GHz}$. Note that only the right half of BZ is shown.}
\label{fig:dispCL} 
\end{figure}

Now we replace the unit cell of the LC circuit with its dual, \textit{i.e.}, a series capacitor and a parallel inductor [see Figure \ref{fig:SchemLCCL}(b)]. As we will see, this circuit shows negative group velocity and has high pass filter characteristics. Writing KVL around the capacitor in unit cell $n$ yields:
\begin{equation}
V_{n-1}-V_{n}=V_{c}\\ \rightarrow L\frac{dI_{n-1}}{dt} - L\frac{dI_n}{dt}=V_c 
\label{eq:CL1}
\end{equation}
The current going through $n$’th capacitor is
\begin{equation}
LC\frac{dV_c}{dt}=I_n + C\frac{d}{dt}(V_n - V_{n+1}) 
\label{eq:CL2}
\end{equation}
By replacing equation (\ref{eq:CL1}) in equation (\ref{eq:CL2}), we get the 2nd order differential form
\begin{equation}
LC\frac{d^2}{dt^2}(I_{n+1}+I_{n-1}-2I_{n})=i_n 
\label{eq:CL3}
\end{equation}
Using the periodic trial solution $I_n=I_0 e^{i(\omega t - kna)}$, the dispersion is found as
\begin{equation}
\omega^2 =  \frac{1}{4LC\sin^2(\frac{ka}{2})}
\label{eq:CL4}
\end{equation}
Figure \ref{fig:dispCL} shows the dispersion of the 1D CL circuit. The negative slope of the dispersion explains why the left-propagating waves are allowed because the group velocity is negative and the sane time the phase velocity which is $\frac{\omega}{k}$ is positive. Different signs of group and phase velocities makes the material or circuit a \textit{left-handed} one. Also, the circuit shows a high pass filter feature as it allows all low $k$ values (small wavelength) frequencies to be propagated. Below a certain frequency, determined by $L$ and $C$ there is no frequency available for all $k$ values. This low cut-off frequency is found in equation (\ref{eq:CL4}) by putting $k_xa= \pm \pi$. For $L = 1~\text{nH}$ and $C = 1~\text{pF}$ we obtain (see Figure \ref{fig:dispCL}),
\begin{equation}
f_{\text{cutoff}}=\frac{1}{4\pi\sqrt{LC}}\approx 2.517~\text{GHz} 
\label{eq:CL5}
\end{equation}

\begin{figure}
\includegraphics[width=3.5in]{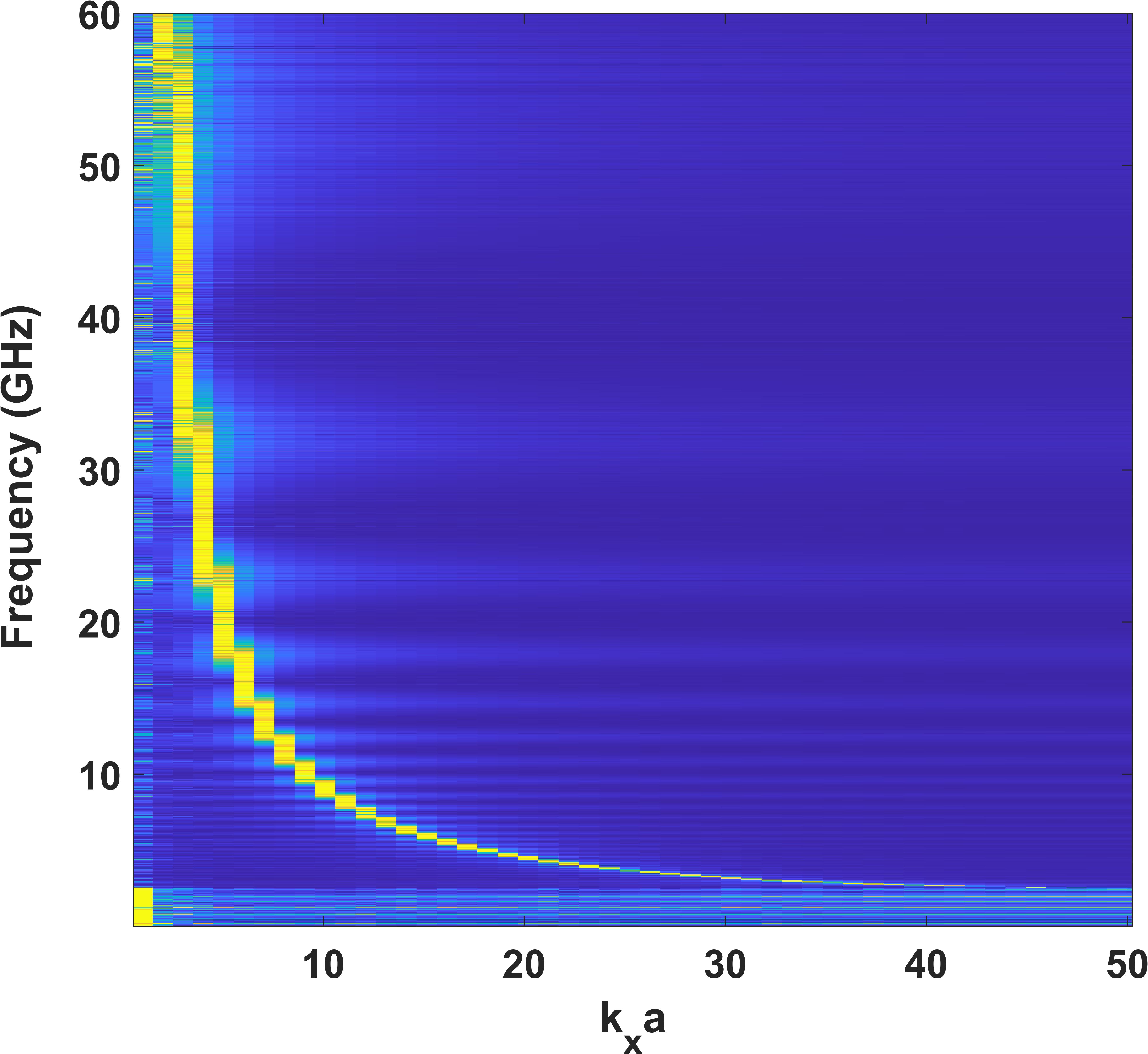}%
\caption{Right half of the 1st BZ found by SED method for the circuit in Figure \ref{fig:SchemLCCL}(b).The low cut-off frequency is about $2~\text{GHz}$. Poor resolution at low $k_x$ values are due to the steep part of the plot, which occupies only one or two $k_x$ pixels. Compare this with Figure \ref{fig:dispCL}.}
\label{fig:dispCLwSED}
\end{figure}

The dispersion calculated by SED is shown in Figure \ref{fig:dispCLwSED}, which closely matches the one found in Figure \ref{fig:dispCL}. The number of unit cells ($k_x$ points) is 100, with the same capacitance and inductance values used in the LC circuit. The noisy data at the bottom shows the stop band below the cut-off frequency (here $2.157~\text{GHz}$). At low $k_x$ values, the branch is steeply rising to infinity, occupying one- or two-pixel areas. By increasing the number of unit cells, the resolution of this part will enhance.

\subsection{\label{sec:level2} A unit cell with two capacitors and two inductors}

\begin{figure}
\includegraphics[width=3.6in]{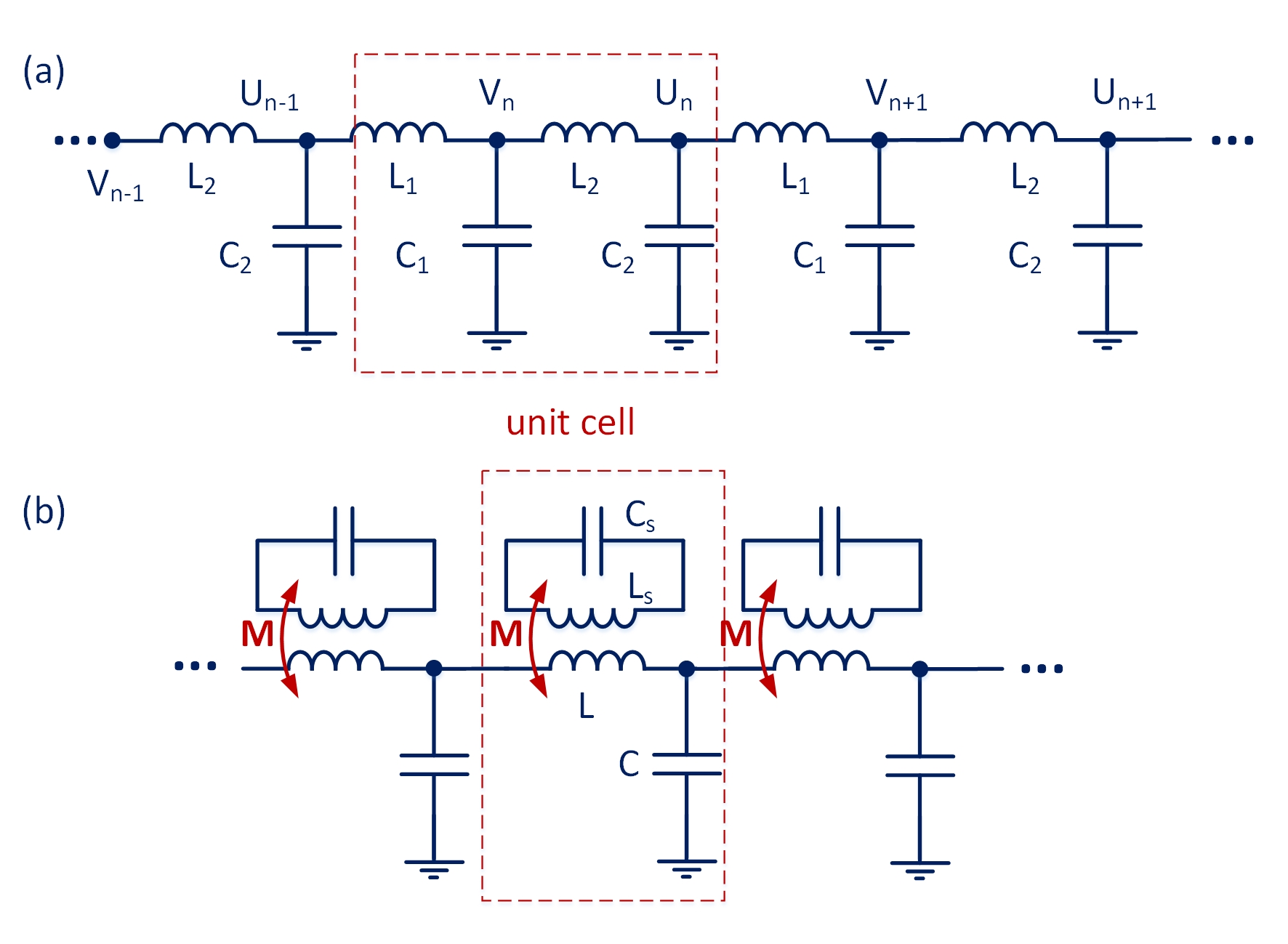}
\caption{(a) The periodic 1D circuit with two different LC sections in each unit cell. (b) A periodic 1D LC circuit with coupling to a resonator of fixed frequency $\omega_s$ in each unit cell.}
\label{fig:Schem2LCAvX}
\end{figure}

\begin{figure}
\includegraphics[width=3.5in]{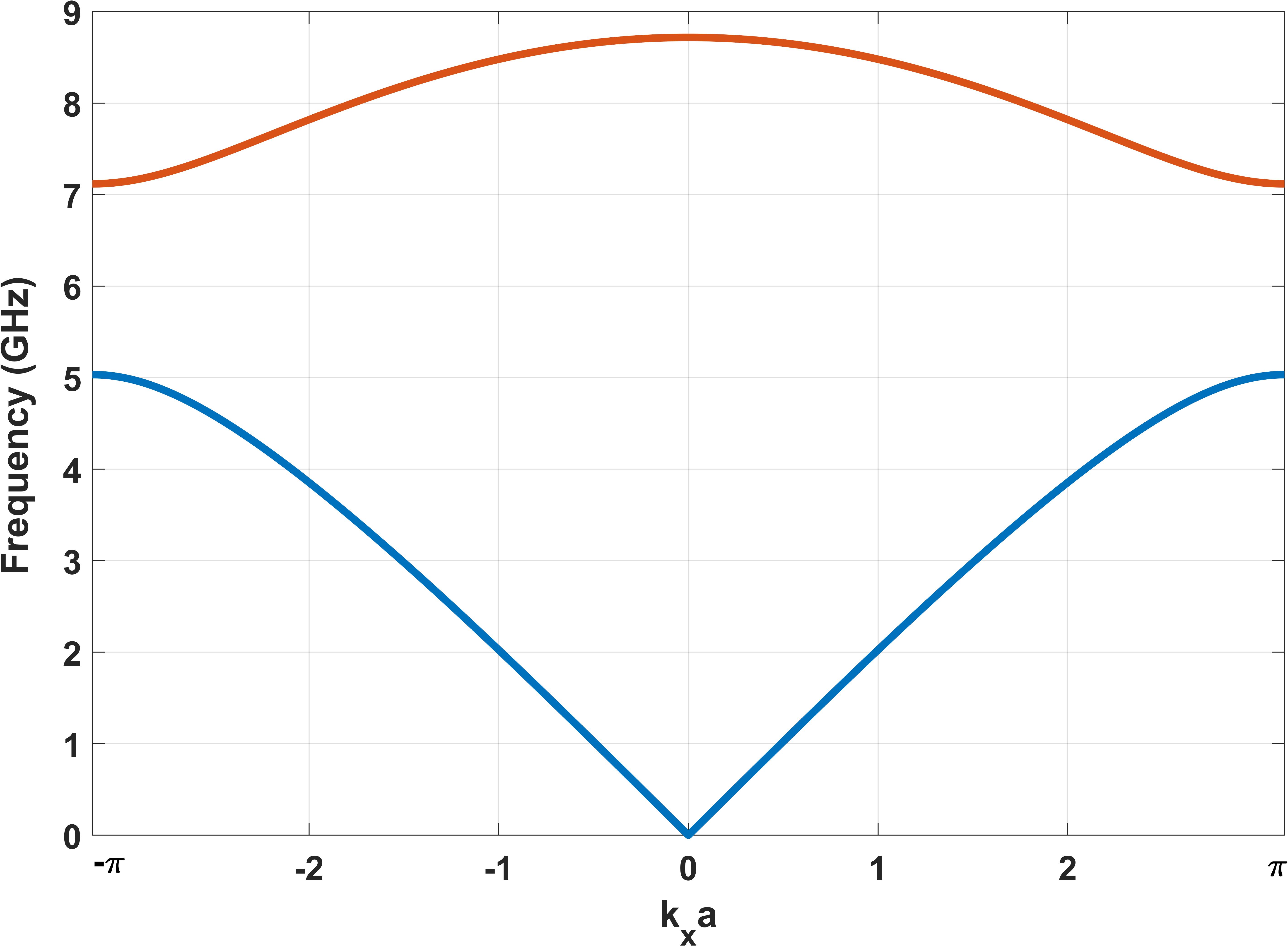}
\caption{The 1st BZ of the periodic circuit with $L = L_1 = L_2 = 1~\text{nH}$ and $C_1 = 1~\text{pF}$ and $C_2 = 2~\text{pF}$ in each unit cell.}
\label{fig:disp2LC}
\end{figure}

Now assume that each unit cell is composed of two inductors and two capacitors of different values, \textit{i.e.}, $L_1$, $L_2$, $C_1$ and $C_2$ as shown in Figure \ref{fig:Schem2LCAvX}(a). For simplicity, let $L = L_1 = L_2$, but we keep $C_1$ and $C_2$ different for the generality of the discussion. The treatment is like the calculation of phonon dispersion for a 1D solid with two different atoms of mass ($m_1$ and $m_2$) which are connected by two springs of stiffness $K_1$ and $K_2$ in the same unit cell \cite{Brilluoin,Kittel}. The only difference is that, instead of the Newton's second law, KVL/KCL is used here to set up two coupled second-order differential equations.
In each unit cell there are two nodes with voltages $U_n$ and $V_n$. For both nodes, we write KCL similar to the case of a simple $LC$ line. For the branch current through the capacitor $C_2$, one obtains
\begin{equation}
\frac{dI_{c2}}{dt}=\frac{d}{dt}(I_{n-1}-I_{n})=\frac{\dot \Phi_{leftofC_2}}{L} - \frac{\dot \Phi_{rightofC_2}}{L} 
\label{eq:2CL1}
\end{equation}
Where left means the inductor on the left side of $C_2$ and right means the inductor on the right side of $C_2$. For $C_1$, one obtains
\begin{equation}
\frac{dI_{c1}}{dt}=\frac{d}{dt}(I_{n}-I_{n+1})=\frac{\dot \Phi_{left of C_1}}{L} - \frac{\dot \Phi_{right of C_1}}{L} 
\label{eq:2CL2}
\end{equation}
Using Lenz’ law and replacing each $d\Phi/dt$ with the voltage difference over the corresponding inductor, and noting that the current going through each capacitor is $I = C dV/dt$, we can write two equations
\begin{eqnarray}
C_1\frac{d^2V_{n}}{dt^2}=\frac{1}{L}(U_{n-1} + U_{n} - 2V_{n}),\\
C_2\frac{d^2U_{n}}{dt^2}=\frac{1}{L}(V_{n} + V_{n+1} - 2U_{n}).
\label{eq:2CL3}
\end{eqnarray}
The above equations are coupled, \textit{i.e.}, the right-hand sides contain both $U$ and $V$, which are unknown. Using the trial solution (Ansatz) for both $U$ and $V$ and replacing them in the above, this set of second-order differential equations are converted to an algebraic equation relating frequency ($\omega$) and weave vector ($k$) as follows. \\
We replace the trail solutions, $U_n=\sum_{k}\tilde U_{k}e^{i(k_xna-\omega t)}$ and $V_n=\sum_{}\tilde V_{k}e^{i(k_xna-\omega t)}$, (which are super positions of plane waves with different wave vectors $k$) in the above equations. Using $\beta_1=1/\sqrt{LC_1}$ and $\beta_2=1/\sqrt{LC_2}$ and factoring out common terms from both sides, we obtain the following matrix equation in which $\tilde U_{k}$ and $\tilde V_{k}$ are unknowns,
\begin{equation}
\label{eqn:matrix2LC}
\begin{bmatrix} \beta_2(1+e^{-ik_xa}) & \omega^2-2\beta_2\\ \omega^2-2\beta_1 & \beta_2(1+e^{+ik_xa}) \end{bmatrix}\cdot\begin{bmatrix}\tilde U_{k}\\ \tilde V_{k}\end{bmatrix}=\boldsymbol{0}=\begin{bmatrix} 0 \\ 0\end{bmatrix}. 
\end{equation}
The above matrix equation is of the form $\boldsymbol{A}\cdot X=0$, hence, to have a non-zero solution, its determinant should be zero [\textit{i.e.}, $\text{det}(\boldsymbol{A})= 0$]. From this condition we have
\begin{equation}
\label{eqn:detzero}
\omega^4 - 2(\beta_1+\beta_2)\omega^2 + 4\sin^2(\frac{k_xa}{2})\beta_1\beta_2 = 0.
\end{equation}
The dispersion is obtained by solving equation (\ref{eqn:detzero}) which gives:
\begin{equation}
\label{eqn:detzerosol}
\omega^2=\frac{C_1+C_2}{LC_1C_2} \pm \frac{1}{LC_1C_2}\sqrt{(C_1+C_2)^2-4C_1C_2\sin^2(\frac{k_xa}{2})}.
\end{equation}
The above dispersion is plotted in Figure \ref{fig:disp2LC}. Recall that this resembles the dispersion of vibrational energies of a 1D periodic solid with two atoms of mass (\textit{e.g.}, $m_1$ and $m_2$) in each unit cell. The factor $(C_1+C_2)/(C_1 C_2)$ in equation (\ref{eqn:detzerosol}) is formally like $(m_1+m_2)/(m_1m_2)$ in the case of phonons. In the latter case, the 1st branch (blue) is called the acoustic, and the second (red) branch is called the optical branch. The modes on $k_xa = \pm \pi$ have zero group velocity, which means they are standing waves in the circuit and do not propagate. The frequencies between $5~\text{GHz}$ and about $7~\text{GHz}$ cannot propagate in the circuit because there is no corresponding wave vector available for them. This frequency band is called the \textit{stopband} or \textit{bandgap}.
The bottom and top edges of the bandgap are found by putting $k_xa = \pm \pi$ in equation (\ref{eqn:detzerosol}),
\begin{equation}
\label{eqn:2LCfHfL}
f_{\text{low}}=\frac{\sqrt{2}}{\pi\sqrt{LC_2}}=5.03~\text{GHz}~,~f_{\text{high}}=\frac{\sqrt{2}}{\pi\sqrt{Lc_1}}=7~\text{GHz}.
\end{equation}
The maximum frequency of the second dispersion branch (cut off) is $8.7~\text{GHz}$ which is found using equation (\ref{eqn:detzerosol}) by putting $k_xa = 0$,
\begin{equation}
\label{eqn:2LCfcutoff}
f_{\text{GHz}}=\frac{\sqrt{2(C_1+C_2)}}{\pi\sqrt{LC_1C_2}}=8.7~\text{GHz}.
\end{equation}
For phonons, this is called the maximum frequency of optical phonons. It is instructive to rewrite the matrix equation (\ref{eqn:matrix2LC}) ($\boldsymbol{A}\cdot X=0$) in a slightly different form as follows
\begin{equation}
\label{eqn:Dynmatrix_1}
\omega^2\begin{bmatrix} 1 & 0\\ 1 & 0 \end{bmatrix}\cdot\begin{bmatrix}\tilde U_{k}\\ \tilde V_{k}\end{bmatrix}=\begin{bmatrix} 2\beta_1 & -\beta_1(1+e^{+ik_xa}) \\ -\beta_2(1+e^{-ik_xa}) & 2\beta_2\end{bmatrix}\begin{bmatrix}\tilde U_{k}\\ \tilde V_{k}\end{bmatrix}. 
\end{equation}
This can be written in the form of an eigenvalue problem as
\begin{equation}
\label{eqn:Dynmatrix_2}
\boldsymbol{D_k}\cdot X = \omega^2 \cdot \boldsymbol{I} \cdot X, 
\end{equation}
where $\boldsymbol{D_k}$ is called \textit{Dynamical Matrix} and $X$ is the eigenvector and is composed of complex wave amplitudes. The eigenvalue,  $\omega^2$, is found by diagonalizing $\boldsymbol{D_k}$. For example, $[\text{amp}, \text{omega2}] = \text{eig}(\boldsymbol{D_k})$ in MATLAB returns the eigenvalues and eigenvectors which correspond to the allowed propagating waves in the circuit. If the Dynamical Matrix is available, it is enough to diagonalize it for each given discrete $k_xa$ within the interval $[0, \pi]$ and plot the eigenvalues for each $k_xa$ value to obtain the same dispersion as in Figure \ref{fig:disp2LC}. 

\begin{figure*}
\centering
{\includegraphics[width=3in]{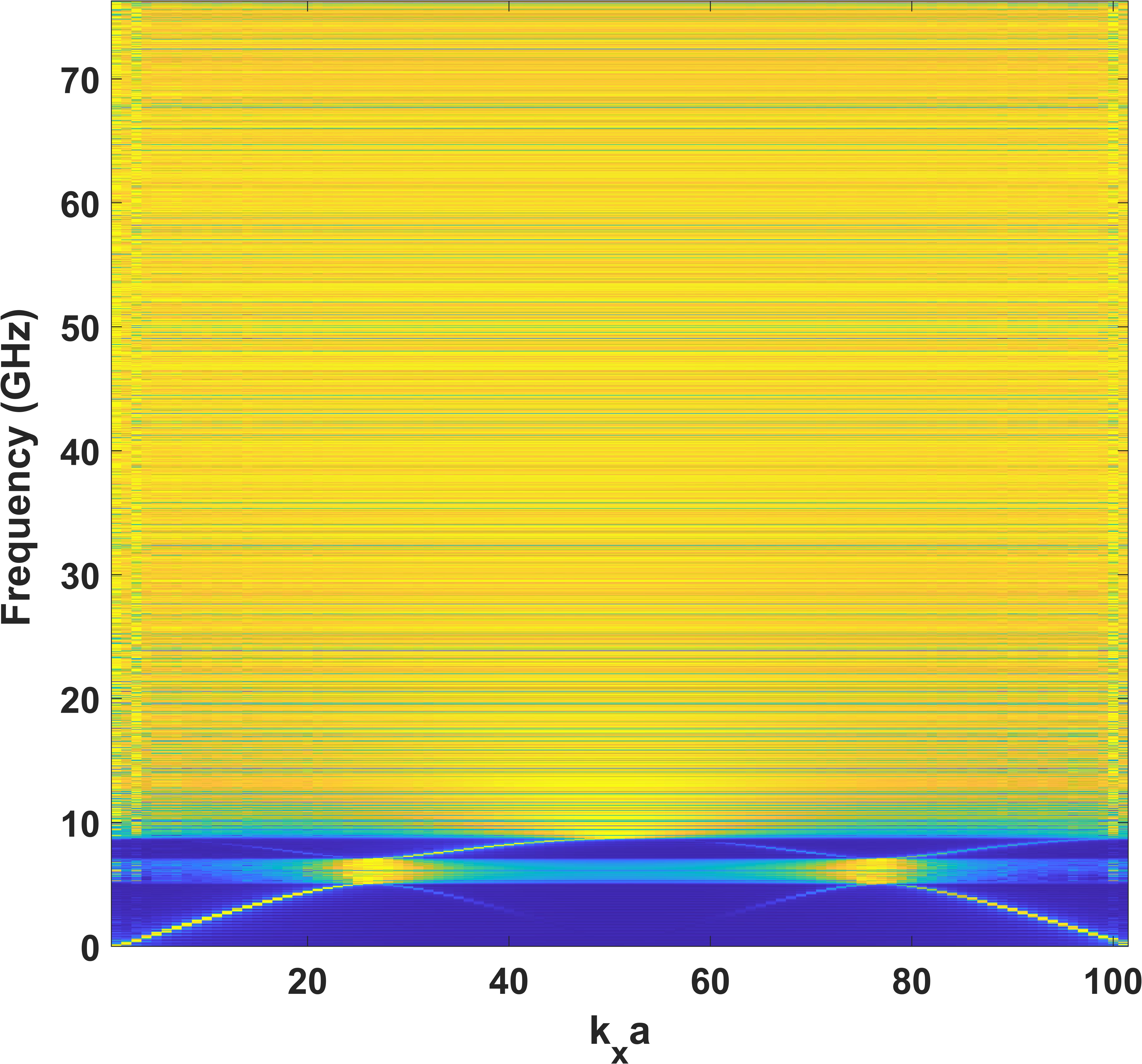}%
\label{fig(a)}}
\hfil
{\includegraphics[width=3in]{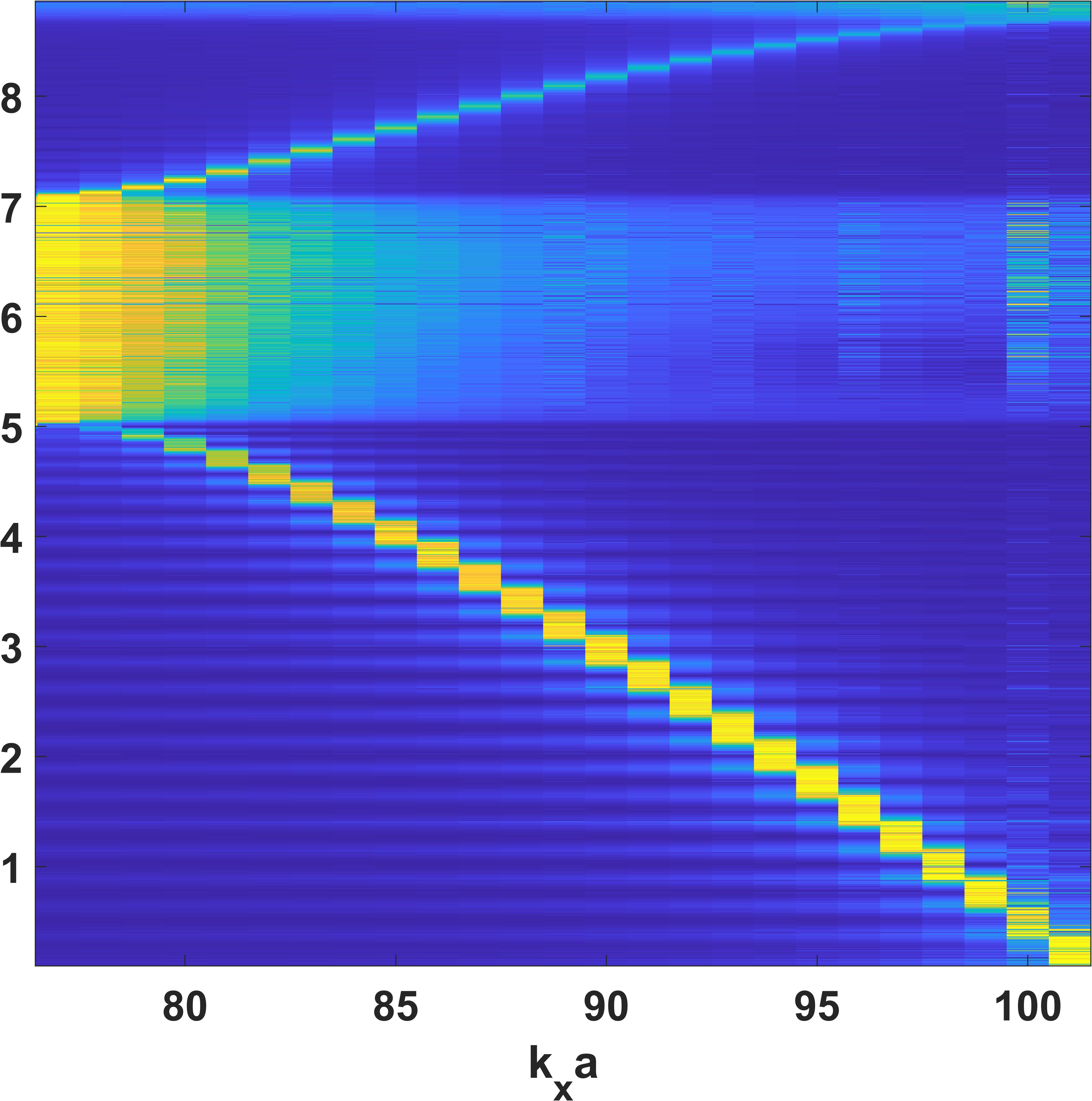}%
\label{fig(b)}}
\caption{(a) The SED for the circuit of Figure \ref{fig:Schem2LCAvX}(a) $L_1 = L_2 = 1~\text{nH}$ and $C_1 = 1~\text{pF}$, $C_2 = 2~\text{pF}$ in the unit cell. (b) One half of BZ cut and magnified from (a), showing the $2~\text{GHz}$ wide bandgap and cut-off frequency of $8.7~\text{GHz}$. Compare this with the left half of Figure \ref{fig:disp2LC}.}
\label{fig:disp2LCsed}
\end{figure*}

Applying the SED method to this circuit results in Figure \ref{fig:disp2LCsed}(a). The highest cut-off frequency is about $8.7~\text{GHz}$, as previously determined by equation (\ref{eqn:2LCfcutoff}) using $L = 1~\text{nH}$ and $C_1 = 1~\text{pF}$ and $C_2 = 2~\text{pF}$. Also there is a bandgap or stop band from $5.03~\text{GHz}$ to $7~\text{GHz}$, similar to the values given by equation (\ref{eqn:2LCfHfL}). Note that since the number of unit cells is 50 this time, the 100-point range of $k_xa$ includes two BZs. Figure \ref{fig:disp2LCsed}(b) shows the 25 $k_x$ points (half of the 1st BZ), which is similar to the left half of the one in Figure \ref{fig:disp2LC}.

\subsection{\label{sec:level2}Coupled Resonators and Avoided Crossing}
Now consider Figure 6(b), where each inductor of the discrete transmission line is mutually coupled to a lumped resonator ring made of $L_s$ and $C_s$. The mutual coupling inductance is $M$. This example helps in teaching the concept of avoided crossing and is relevant to the case of time-independent perturbation with degeneracy. In this case, we show how, at a certain $k_x$ point, there are two degenerate and equal eigen frequencies ($5~\text{GHz}$), and this degeneracy is lifted by the amount, determined by the mutual coupling constant $M$. It is also helpful to show students what kind of structure leads to a flat line in the dispersion, \textit{i.e.}, a constant frequency (energy) line for all values of $k$. In our case, a lumped (localized) resonator leads to the constant $5~\text{GHz}$ line. 
To find the dispersion of a $LC$ line coupled with the resonator, we first write the KCL in the $n$’th node. The currents and voltages in the $LC$ line are represented by $i_n$ and $v_n$. The corresponding quantities for the $n$’th $L_sC_s$ loop are shown by adding a tilde on top, \textit{e.g.}, $\tilde i_n$ and $\tilde U _n$. The current entering the $n$'th unit cell from the previous one is divided between the capacitor, $C$, and the inductor, $L$, \textit{i.e.} (see Figure \ref{fig:Schem2LCAvX}(b)),  
\begin{equation}
\label{eqn:AvX1}
i_{n-1}=i_{n}+C\frac{dv_n}{dt}.
\end{equation}
After differentiating both sides with respect to time, we have
\begin{equation}
\label{eqn:AvX2}
\frac{di_{n-1}}{dt}=\frac{di_n}{dt}+C\frac{d^2v_n}{dt^2}. 
\end{equation}
For each unit cell, due to the mutual inductance, the following voltage relations for ($n-1$) and $n$ unit cells hold
\begin{eqnarray}
\label{eqn:AvX3}
\frac{di_{n-1}}{dt}=v_{n-1}-v_{n} - M\frac{d\tilde i_{n-1}}{dt}, \label{sub:1}\\
\frac{di_{n}}{dt}=v_{n}-v_{n+1} - M\frac{d\tilde i_{n}}{dt}.\label{sub:2}
\end{eqnarray}
Inserting the above in equation (\ref{eqn:AvX2}), results in
\begin{equation}
\label{eqn:AvX4}
v_{n+1}+v_{n-1}-2v_{n}-LC\frac{d^2v_{n}}{dt^2}=M\frac{d\tilde i_{n-1}}{dt}-M\frac{d\tilde i_n}{dt}.
\end{equation}
To write $\tilde i_{n}$ in terms of $\tilde U_n$, note that the current inside each $L_sC_s$ loop is found from the capacitor voltage which is $\tilde i_n=-C_s\frac{d\tilde U_n}{dt}$. By replacing this in (\ref{sub:2}), we have
\begin{equation}
\label{eqn:AvX5}
v_{n+1}+v_{n-1}-2v_{n}-LC\frac{d^2v_{n}}{dt^2}=-MC_s\frac{d^2\tilde U_{n-1}}{dt^2}+MC_s\frac{d^2\tilde U_n}{dt^2}.
\end{equation}
Equation (\ref{eqn:AvX5}) is the first one required to find the dispersion of the modes, and it is written in terms of two voltages, $v$ and $U$. Now, a $KVL$ is written inside the $n$’th $L_sC_s$ loop using $\tilde i_n=-C_s\frac{d\tilde U_n}{dt}$
\begin{equation}
\label{eqn:AvX6}
L_s\frac{d\tilde i_n}{dt} + M\frac{di_n}{dt}=\tilde U_n. 
\end{equation}
Thereafter, $di_n/dt$ is replaced with its equivalent given in equation (\ref{sub:2}) and $d\tilde i_n/dt$ is replaced with the help of equation (\ref{eqn:AvX4}). For brevity, the dot notation is used for differentiation with respect to time, \textit{i.e.}, $dv/dt = \dot v$. The new form of equation (\ref{eqn:AvX6}) will be
\begin{equation}
\label{eqn:AvX7}
\frac{M}{L}(v_n -v_{n+1}) + \frac{M^2C_s}{L} \ddot U_n - C_sL_s\ddot U_n - U_n =0. 
\end{equation}
Equations (\ref{eqn:AvX5}) and (\ref{eqn:AvX7}) are grouped in a matrix form. 
\begin{equation}
\label{eqn:AvX8}
\left\{
\begin{array}{cc}
v_{n+1} + v_{n-1} -2v_{n}-LC\ddot v_{n} + MC_s(\ddot U_{n-1}-\ddot U_{n})=0 \\
\frac{M}{L}(v_{n}-v_{n+1}) +\frac{M^2C_s}{L}\ddot U_{n} - L_sC_s\ddot U_{n} - U_{n}=0.
\end{array} \right.
\end{equation}
By substituting the trial solutions like $U_n=\sum_k U_k e^{i(k_xna-\omega t)}$ and $V_n=\sum_k V_k e^{i(k_xna-\omega t)}$ in the above, the following matrix equation is formed,
\begin{equation}
\label{eqn:AvX9}
\begin{bmatrix} 2(e^{+ik_xa}-1)+LC\omega^2 & MC_s\omega^2(1-e^{ik_xa})\\ \frac{M}{L}(1-e^{-ik_xa}) & -\frac{M^2C_s\omega^2}{L}+L_sC_s\omega^2-1\end{bmatrix}\begin{bmatrix}\tilde V_{k}\\ \tilde U_{k}\end{bmatrix}=\begin{bmatrix}0\\ 0\end{bmatrix}.
\end{equation}
To have a non-zero solution, the determinant of the above matrix is must be zero. From this and using the shorthand notations $\omega_0=1/\sqrt{LC}$ and $\Omega=1/\sqrt{L_sC_s}$, the dispersion relation is extracted.
\begin{equation}
\label{eqn:AvX11}
\cos(k_xa) = 1 - \frac{LC}{2}\omega^2 + \frac{\omega^4}{\omega^2-\Omega^2}(\frac{M^2C}{2L_s}). 
\end{equation}

\begin{figure}
\includegraphics[width=3.5in]{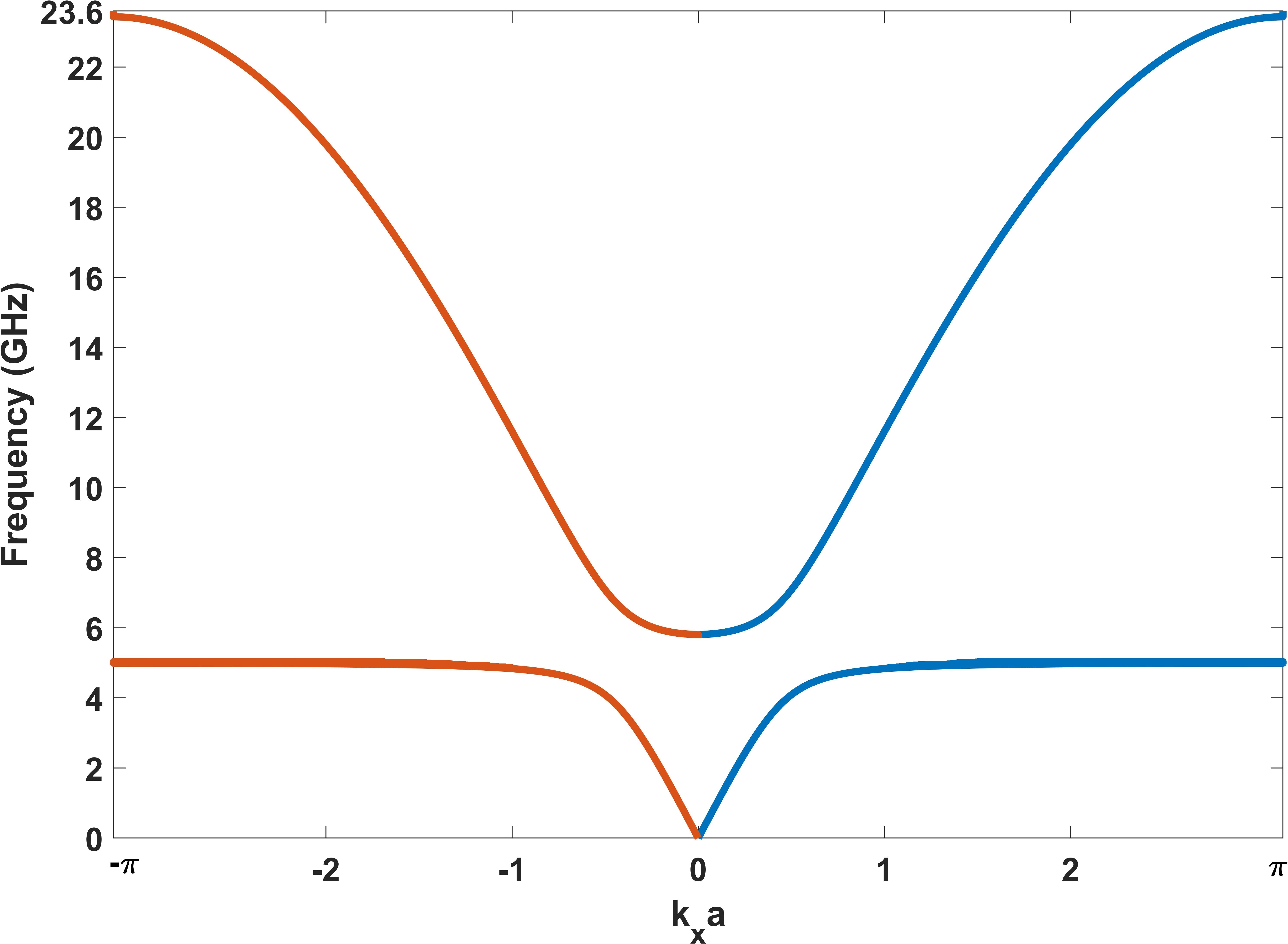}
\caption{Avoided crossing in the dispersion plot of a LC circuit interacting with a ring resonator of $f_s = 5~\text{GHz}$ in each unit cell as shown in Figure \ref{fig:Schem2LCAvX}(b) . The coupling (interaction) causes a gap formation from $5.035~\text{GHz}$ to $5.812~\text{GHz}$.}
\label{fig:AvXmatlab}
\end{figure}

Note that in contrast to equations (\ref{eq:LC4}), (\ref{eq:CL4}), and (\ref{eqn:detzerosol}), where $\omega$ was a function of $k_x$, here $k_x$ must be found numerically by sweeping the frequency, $\omega$.\\
With the example values of $L_s = 1~\text{nH}$, $C_s =1~\text{pF}$, $L = 1~\text{nH}$, and $C=0.25~\text{pF}$; the resulting dispersion is plotted in Figure \ref{fig:AvXmatlab}. It is shown that the bandgap (stopband) begins at $5.035~\text{GHz}$ and ends at $5.812~\text{GHz}$. Note that if coupling did not exist, \textit{i.e.}, $M=0$, we would have the same dispersion as the one for $LC$ circuit (as in Figure \ref{fig:dispLC}) plus a horizontal line at $5~\text{GHz}$ for all values of $k_x$. It is because the resonance frequency of the ring resonator is always $\Omega/2\pi=1/(2\pi\sqrt{L_sC_s})=5~\text{GHz}$, and the localized oscillation on this resonator needs all $k_x$ (wavelength) values in order to get formed. This resembles the physics of a defect in a crystal on which an electron is localized or trapped, and as a result, a constant-energy line is observed in the dispersion. 

\begin{figure}
\includegraphics[width=3.5in]{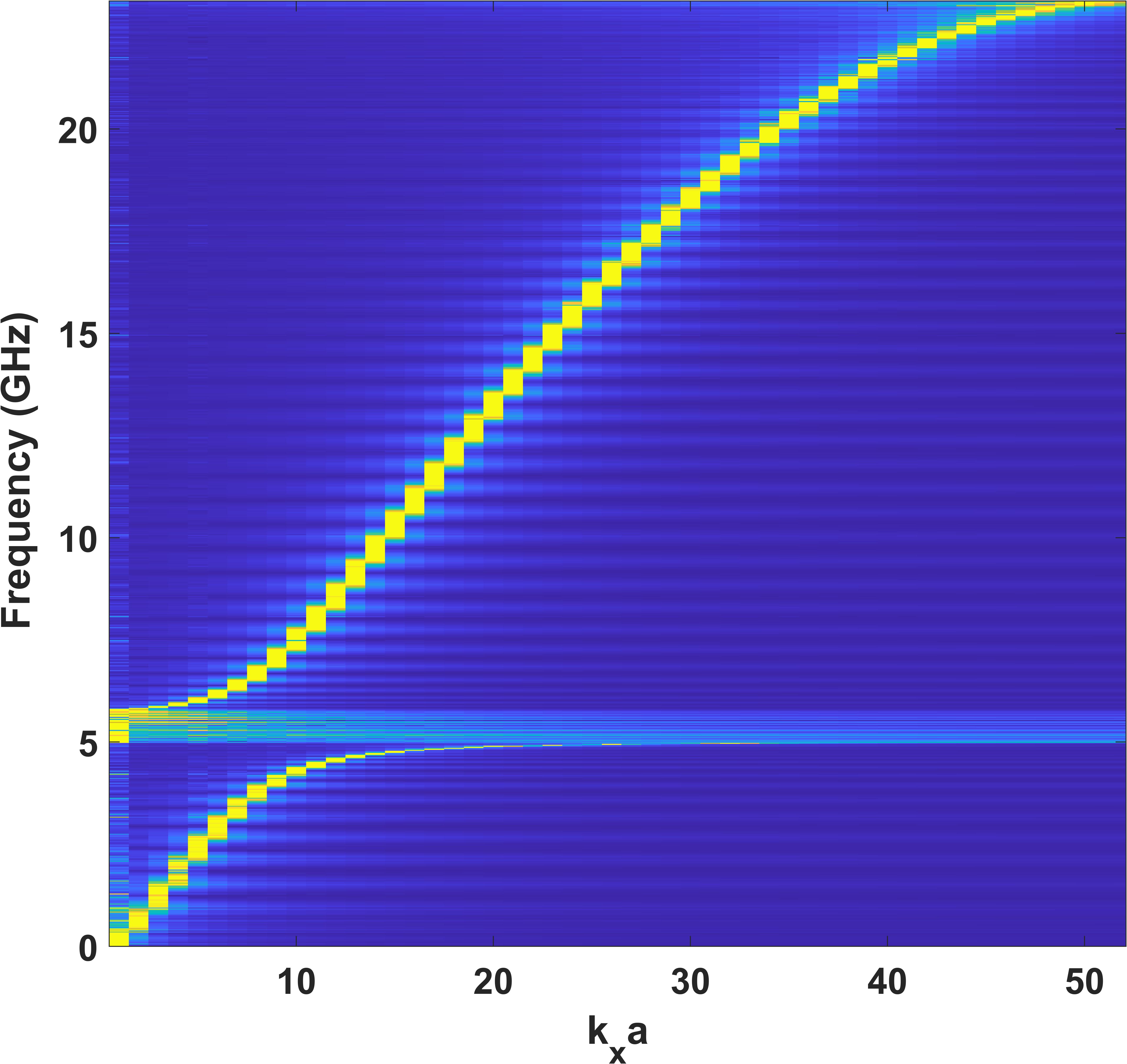}
\caption{The SED for LC circuit coupled to ring resonator in each unit cell (\ref{fig:Schem2LCAvX}(b)) for the values of inductance and capacitance given in the text. Compare this with the right half of Figure \ref{fig:AvXmatlab}.}
\label{fig:dispAvXSED}
\end{figure}

By turning on the coupling (here $M = 0.5$), the dispersion plots on the overlapping point (same frequency of $5~\text{GHz}$) split into two branches of different frequencies. This is what is called \textit{avoided crossing}. The SED method returns Figure \ref{fig:dispAvXSED} where the 1st half of the BZ is plotted and it is similar to the one obtained by the analytic method in Figure \ref{fig:AvXmatlab}. Note the resemblance of avoided crossing (noisy part of the figure) and the highest cut-off frequency, which is $23.6~\text{GHz}$ in both methods. 

\section{\label{sec:level1} Conclusions}
We demonstrated the implementation of the spectral energy density (SED) method using the circuit analog quantities for phonons. From that, we obtained dispersion of four different periodic 1D circuits based on discrete unit cells composed of L and C. While the examples presented in this work are trivial and circuit analysis seems to be the easier way to get the dispersion, firstly, the proposed approach is helpful in increasing the student's grasp of the concepts like BZ, dispersion, zone-folding, and dynamical matrix. Secondly, in the case of a periodic circuit of a complex or nonlinear topology for which the standard KVL/KCL-based method is not straightforward, the SED approach proves helpful. Thirdly, the presented approach can be used in emulating the thermal transport physics of nanomaterials, \textit{e.g.}, heat rectifiers \cite{Maznev2014,Shiri2019} using commercial circuit simulators and translating the language of mass and spring to that of inductance and capacitance.


\begin{acknowledgments}
Authors acknowledge the instructive comments given by Dr. M. P. Anantram and Dr. Anita Fadavi Roudsari.
\end{acknowledgments}

\section{Data Availability}
The data that support the findings of this study are available from the corresponding author upon reasonable request.

%
\section{References}
\nocite{*}
\bibliography{Manuscript_JAP_Shiri_Isacsson_SED_Nov23}

\providecommand{\noopsort}[1]{}\providecommand{\singleletter}[1]{#1}%
\begin{thebibliography}{24}%
\makeatletter
\providecommand \@ifxundefined [1]{%
 \@ifx{#1\undefined}
}%
\providecommand \@ifnum [1]{%
 \ifnum #1\expandafter \@firstoftwo
 \else \expandafter \@secondoftwo
 \fi
}%
\providecommand \@ifx [1]{%
 \ifx #1\expandafter \@firstoftwo
 \else \expandafter \@secondoftwo
 \fi
}%
\providecommand \natexlab [1]{#1}%
\providecommand \enquote  [1]{``#1''}%
\providecommand \bibnamefont  [1]{#1}%
\providecommand \bibfnamefont [1]{#1}%
\providecommand \citenamefont [1]{#1}%
\providecommand \href@noop [0]{\@secondoftwo}%
\providecommand \href [0]{\begingroup \@sanitize@url \@href}%
\providecommand \@href[1]{\@@startlink{#1}\@@href}%
\providecommand \@@href[1]{\endgroup#1\@@endlink}%
\providecommand \@sanitize@url [0]{\catcode `\\12\catcode `\$12\catcode
  `\&12\catcode `\#12\catcode `\^12\catcode `\_12\catcode `\%12\relax}%
\providecommand \@@startlink[1]{}%
\providecommand \@@endlink[0]{}%
\providecommand \url  [0]{\begingroup\@sanitize@url \@url }%
\providecommand \@url [1]{\endgroup\@href {#1}{\urlprefix }}%
\providecommand \urlprefix  [0]{URL }%
\providecommand \Eprint [0]{\href }%
\providecommand \doibase [0]{http://dx.doi.org/}%
\providecommand \selectlanguage [0]{\@gobble}%
\providecommand \bibinfo  [0]{\@secondoftwo}%
\providecommand \bibfield  [0]{\@secondoftwo}%
\providecommand \translation [1]{[#1]}%
\providecommand \BibitemOpen [0]{}%
\providecommand \bibitemStop [0]{}%
\providecommand \bibitemNoStop [0]{.\EOS\space}%
\providecommand \EOS [0]{\spacefactor3000\relax}%
\providecommand \BibitemShut  [1]{\csname bibitem#1\endcsname}%
\let\auto@bib@innerbib\@empty
\bibitem [{\citenamefont {Brilluoin}(1953)}]{Brilluoin}%
  \BibitemOpen
  \bibfield  {author} {\bibinfo {author} {\bibfnamefont {L.}~\bibnamefont
  {Brilluoin}},\ }in\ \href@noop {} {\emph {\bibinfo {booktitle} {Wave
  Propagation in Periodic Structures}}},\ \bibinfo {series and number}
  {\bibinfo {number} {Dover Publications}}\ (\bibinfo {year}
  {1953})\BibitemShut {NoStop}%
\bibitem [{\citenamefont {Kron}(1959)}]{Kron}%
  \BibitemOpen
  \bibfield  {author} {\bibinfo {author} {\bibfnamefont {G.}~\bibnamefont
  {Kron}},\ }in\ \href@noop {} {\emph {\bibinfo {booktitle} {Tensors for
  circuits}}},\ \bibinfo {series and number} {\bibinfo {number} {Dover
  Publications}}\ (\bibinfo {year} {1959})\BibitemShut {NoStop}%
\bibitem [{\citenamefont {Shockley}(1950)}]{Shockley}%
  \BibitemOpen
  \bibfield  {author} {\bibinfo {author} {\bibfnamefont {W.}~\bibnamefont
  {Shockley}},\ }in\ \href@noop {} {\emph {\bibinfo {booktitle} {Electrons and
  Holes in Semiconductors: With Applications to Transistor Electronics}}},\
  \bibinfo {series and number} {\bibinfo {number} {Van Nostrand}}\ (\bibinfo
  {year} {1950})\BibitemShut {NoStop}%
\bibitem [{\citenamefont {Toda}(1989)}]{Toda}%
  \BibitemOpen
  \bibfield  {author} {\bibinfo {author} {\bibfnamefont {M.}~\bibnamefont
  {Toda}},\ }in\ \href@noop {} {\emph {\bibinfo {booktitle} {Theory of
  Nonlinear Lattices}}},\ \bibinfo {series and number} {\bibinfo {number}
  {Springer-Verlag Berlin Heidelberg}}\ (\bibinfo {year} {1989})\BibitemShut
  {NoStop}%
\bibitem [{\citenamefont {Giambo}, \citenamefont {Pantano},\ and\ \citenamefont
  {Tucci}(1984)}]{Giambo1984}%
  \BibitemOpen
  \bibfield  {author} {\bibinfo {author} {\bibfnamefont {S.}~\bibnamefont
  {Giambo}}, \bibinfo {author} {\bibfnamefont {P.}~\bibnamefont {Pantano}}, \
  and\ \bibinfo {author} {\bibfnamefont {P.}~\bibnamefont {Tucci}},\ }\bibfield
   {title} {\enquote {\bibinfo {title} {An electrical model for the korteweg-de
  vries equation},}\ }\href {\doibase 10.1119/1.13685} {\bibfield  {journal}
  {\bibinfo  {journal} {American Journal of Physics}\ }\textbf {\bibinfo
  {volume} {52}},\ \bibinfo {pages} {238--243} (\bibinfo {year}
  {1984})}\BibitemShut {NoStop}%
\bibitem [{\citenamefont {Price}(2022)}]{Price2022}%
  \BibitemOpen
  \bibfield  {author} {\bibinfo {author} {\bibfnamefont {H.}~\bibnamefont
  {Price}},\ }\bibfield  {title} {\enquote {\bibinfo {title} {Simulating
  four-dimensional physics in the laboratory},}\ }\href {\doibase
  10.1063/PT.3.4981} {\bibfield  {journal} {\bibinfo  {journal} {Physics
  Today}\ }\textbf {\bibinfo {volume} {75}},\ \bibinfo {pages} {38--44}
  (\bibinfo {year} {2022})}\BibitemShut {NoStop}%
\bibitem [{\citenamefont {Dong}, \citenamefont {Juri\ifmmode \check{c}\else
  \v{c}\fi{}i\ifmmode~\acute{c}\else \'{c}\fi{}},\ and\ \citenamefont
  {Roy}(2021)}]{Dong2021}%
  \BibitemOpen
  \bibfield  {author} {\bibinfo {author} {\bibfnamefont {J.}~\bibnamefont
  {Dong}}, \bibinfo {author} {\bibfnamefont {V.}~\bibnamefont {Juri\ifmmode
  \check{c}\else \v{c}\fi{}i\ifmmode~\acute{c}\else \'{c}\fi{}}}, \ and\
  \bibinfo {author} {\bibfnamefont {B.}~\bibnamefont {Roy}},\ }\bibfield
  {title} {\enquote {\bibinfo {title} {Topolectric circuits: Theory and
  construction},}\ }\href {\doibase 10.1103/PhysRevResearch.3.023056}
  {\bibfield  {journal} {\bibinfo  {journal} {Phys. Rev. Research}\ }\textbf
  {\bibinfo {volume} {3}},\ \bibinfo {pages} {023056} (\bibinfo {year}
  {2021})}\BibitemShut {NoStop}%
\bibitem [{\citenamefont {Lee}\ \emph {et~al.}(2018)\citenamefont {Lee},
  \citenamefont {Imhof}, \citenamefont {Berger}, \citenamefont {Bayer},
  \citenamefont {Brehm}, \citenamefont {Molenkamp}, \citenamefont {Kiessling},\
  and\ \citenamefont {Thomale}}]{Lee2018}%
  \BibitemOpen
  \bibfield  {author} {\bibinfo {author} {\bibfnamefont {C.~H.}\ \bibnamefont
  {Lee}}, \bibinfo {author} {\bibfnamefont {S.}~\bibnamefont {Imhof}}, \bibinfo
  {author} {\bibfnamefont {C.}~\bibnamefont {Berger}}, \bibinfo {author}
  {\bibfnamefont {F.}~\bibnamefont {Bayer}}, \bibinfo {author} {\bibfnamefont
  {J.}~\bibnamefont {Brehm}}, \bibinfo {author} {\bibfnamefont {L.~W.}\
  \bibnamefont {Molenkamp}}, \bibinfo {author} {\bibfnamefont {T.}~\bibnamefont
  {Kiessling}}, \ and\ \bibinfo {author} {\bibfnamefont {R.}~\bibnamefont
  {Thomale}},\ }\bibfield  {title} {\enquote {\bibinfo {title} {Topolectrical
  circuits},}\ }\href {\doibase 10.1038/s42005-018-0035-2} {\bibfield
  {journal} {\bibinfo  {journal} {Communications Physics}\ }\textbf {\bibinfo
  {volume} {1}},\ \bibinfo {pages} {39} (\bibinfo {year} {2018})}\BibitemShut
  {NoStop}%
\bibitem [{\citenamefont {Thomas}\ \emph {et~al.}(2010)\citenamefont {Thomas},
  \citenamefont {Turney}, \citenamefont {Iutzi}, \citenamefont {Amon},\ and\
  \citenamefont {McGaughey}}]{Thomas2010}%
  \BibitemOpen
  \bibfield  {author} {\bibinfo {author} {\bibfnamefont {J.~A.}\ \bibnamefont
  {Thomas}}, \bibinfo {author} {\bibfnamefont {J.~E.}\ \bibnamefont {Turney}},
  \bibinfo {author} {\bibfnamefont {R.~M.}\ \bibnamefont {Iutzi}}, \bibinfo
  {author} {\bibfnamefont {C.~H.}\ \bibnamefont {Amon}}, \ and\ \bibinfo
  {author} {\bibfnamefont {A.~J.~H.}\ \bibnamefont {McGaughey}},\ }\bibfield
  {title} {\enquote {\bibinfo {title} {Predicting phonon dispersion relations
  and lifetimes from the spectral energy density},}\ }\href {\doibase
  10.1103/PhysRevB.81.081411} {\bibfield  {journal} {\bibinfo  {journal} {Phys.
  Rev. B}\ }\textbf {\bibinfo {volume} {81}},\ \bibinfo {pages} {081411}
  (\bibinfo {year} {2010})}\BibitemShut {NoStop}%
\bibitem [{\citenamefont {Larkin}\ \emph {et~al.}(2014)\citenamefont {Larkin},
  \citenamefont {Turney}, \citenamefont {Massicotte}, \citenamefont {Amon},\
  and\ \citenamefont {McGaughey}}]{Larkin2014}%
  \BibitemOpen
  \bibfield  {author} {\bibinfo {author} {\bibfnamefont {J.~M.}\ \bibnamefont
  {Larkin}}, \bibinfo {author} {\bibfnamefont {J.~E.}\ \bibnamefont {Turney}},
  \bibinfo {author} {\bibfnamefont {A.~D.}\ \bibnamefont {Massicotte}},
  \bibinfo {author} {\bibfnamefont {C.~H.}\ \bibnamefont {Amon}}, \ and\
  \bibinfo {author} {\bibfnamefont {A.~J.~H.}\ \bibnamefont {McGaughey}},\
  }\bibfield  {title} {\enquote {\bibinfo {title} {Comparison and evaluation of
  spectral energy methods for predicting phonon properties},}\ }\href {\doibase
  doi:10.1166/jctn.2014.3345} {\bibfield  {journal} {\bibinfo  {journal}
  {Journal of Computational and Theoretical Nanoscience}\ }\textbf {\bibinfo
  {volume} {11}},\ \bibinfo {pages} {249--256} (\bibinfo {year}
  {2014})}\BibitemShut {NoStop}%
\bibitem [{\citenamefont {Thompson}\ \emph {et~al.}(2022)\citenamefont
  {Thompson}, \citenamefont {Aktulga}, \citenamefont {Berger}, \citenamefont
  {Bolintineanu}, \citenamefont {Brown}, \citenamefont {Crozier}, \citenamefont
  {{in 't Veld}}, \citenamefont {Kohlmeyer}, \citenamefont {Moore},
  \citenamefont {Nguyen}, \citenamefont {Shan}, \citenamefont {Stevens},
  \citenamefont {Tranchida}, \citenamefont {Trott},\ and\ \citenamefont
  {Plimpton}}]{THOMPSON2022}%
  \BibitemOpen
  \bibfield  {author} {\bibinfo {author} {\bibfnamefont {A.~P.}\ \bibnamefont
  {Thompson}}, \bibinfo {author} {\bibfnamefont {H.~M.}\ \bibnamefont
  {Aktulga}}, \bibinfo {author} {\bibfnamefont {R.}~\bibnamefont {Berger}},
  \bibinfo {author} {\bibfnamefont {D.~S.}\ \bibnamefont {Bolintineanu}},
  \bibinfo {author} {\bibfnamefont {W.~M.}\ \bibnamefont {Brown}}, \bibinfo
  {author} {\bibfnamefont {P.~S.}\ \bibnamefont {Crozier}}, \bibinfo {author}
  {\bibfnamefont {P.~J.}\ \bibnamefont {{in 't Veld}}}, \bibinfo {author}
  {\bibfnamefont {A.}~\bibnamefont {Kohlmeyer}}, \bibinfo {author}
  {\bibfnamefont {S.~G.}\ \bibnamefont {Moore}}, \bibinfo {author}
  {\bibfnamefont {T.~D.}\ \bibnamefont {Nguyen}}, \bibinfo {author}
  {\bibfnamefont {R.}~\bibnamefont {Shan}}, \bibinfo {author} {\bibfnamefont
  {M.~J.}\ \bibnamefont {Stevens}}, \bibinfo {author} {\bibfnamefont
  {J.}~\bibnamefont {Tranchida}}, \bibinfo {author} {\bibfnamefont
  {C.}~\bibnamefont {Trott}}, \ and\ \bibinfo {author} {\bibfnamefont {S.~J.}\
  \bibnamefont {Plimpton}},\ }\bibfield  {title} {\enquote {\bibinfo {title}
  {{LAMMPS} - a flexible simulation tool for particle-based materials modeling
  at the atomic, meso, and continuum scales},}\ }\href {\doibase
  https://doi.org/10.1016/j.cpc.2021.108171} {\bibfield  {journal} {\bibinfo
  {journal} {Computer Physics Communications}\ }\textbf {\bibinfo {volume}
  {271}},\ \bibinfo {pages} {108171} (\bibinfo {year} {2022})}\BibitemShut
  {NoStop}%
\bibitem [{\citenamefont {Gale}\ and\ \citenamefont {Rohl}(2003)}]{Gulp2003}%
  \BibitemOpen
  \bibfield  {author} {\bibinfo {author} {\bibfnamefont {J.~D.}\ \bibnamefont
  {Gale}}\ and\ \bibinfo {author} {\bibfnamefont {A.~L.}\ \bibnamefont
  {Rohl}},\ }\bibfield  {title} {\enquote {\bibinfo {title} {The general
  utility lattice program ({GULP})},}\ }\href {\doibase
  10.1080/0892702031000104887} {\bibfield  {journal} {\bibinfo  {journal}
  {Molecular Simulation}\ }\textbf {\bibinfo {volume} {29}},\ \bibinfo {pages}
  {291--341} (\bibinfo {year} {2003})}\BibitemShut {NoStop}%
\bibitem [{\citenamefont {Zou}(2008)}]{Zou2008}%
  \BibitemOpen
  \bibfield  {author} {\bibinfo {author} {\bibfnamefont {J.}~\bibnamefont
  {Zou}},\ }\bibfield  {title} {\enquote {\bibinfo {title} {Calculation of
  phonon dispersion in semiconductor nanostructures: An undergraduate
  computational project},}\ }\href {\doibase 10.1119/1.2825396} {\bibfield
  {journal} {\bibinfo  {journal} {American Journal of Physics}\ }\textbf
  {\bibinfo {volume} {76}},\ \bibinfo {pages} {460--463} (\bibinfo {year}
  {2008})}\BibitemShut {NoStop}%
\bibitem [{\citenamefont {Balandin}(2007)}]{Balandin2007}%
  \BibitemOpen
  \bibfield  {author} {\bibinfo {author} {\bibfnamefont {A.~A.}\ \bibnamefont
  {Balandin}},\ }\bibfield  {title} {\enquote {\bibinfo {title} {Nanophononics:
  fine-tuning phonon dispersion in semiconductor nanostructures},}\ }\href
  {http://inis.iaea.org/search/search.aspx?orig_q=RN:40072087} {\bibfield
  {journal} {\bibinfo  {journal} {Moldavian Journal of the Physical Sciences}\
  }\textbf {\bibinfo {volume} {6}},\ \bibinfo {pages} {32--38} (\bibinfo {year}
  {2007})}\BibitemShut {NoStop}%
\bibitem [{\citenamefont {Aumentado}(2020)}]{Aumentado2020}%
  \BibitemOpen
  \bibfield  {author} {\bibinfo {author} {\bibfnamefont {J.}~\bibnamefont
  {Aumentado}},\ }\bibfield  {title} {\enquote {\bibinfo {title}
  {Superconducting parametric amplifiers: The state of the art in josephson
  parametric amplifiers},}\ }\href {\doibase 10.1109/MMM.2020.2993476}
  {\bibfield  {journal} {\bibinfo  {journal} {IEEE Microwave Magazine}\
  }\textbf {\bibinfo {volume} {21}},\ \bibinfo {pages} {45--59} (\bibinfo
  {year} {2020})}\BibitemShut {NoStop}%
\bibitem [{\citenamefont {{W}hiteley~{R}esearch
  {I}ncorporated}(2022)}]{WRSpice}%
  \BibitemOpen
  \bibfield  {author} {\bibinfo {author} {\bibnamefont {{W}hiteley~{R}esearch
  {I}ncorporated}},\ }\href {http://www.wrcad.com/wrspice.html} {\enquote
  {\bibinfo {title} {W{R}spice, release 4.3.17},}\ } (\bibinfo {year}
  {2022})\BibitemShut {NoStop}%
\bibitem [{\citenamefont {Martin}\ \emph {et~al.}(2003)\citenamefont {Martin},
  \citenamefont {Falcone}, \citenamefont {Bonache}, \citenamefont {Marques},\
  and\ \citenamefont {Sorolla}}]{Martin2003}%
  \BibitemOpen
  \bibfield  {author} {\bibinfo {author} {\bibfnamefont {F.}~\bibnamefont
  {Martin}}, \bibinfo {author} {\bibfnamefont {F.}~\bibnamefont {Falcone}},
  \bibinfo {author} {\bibfnamefont {J.}~\bibnamefont {Bonache}}, \bibinfo
  {author} {\bibfnamefont {R.}~\bibnamefont {Marques}}, \ and\ \bibinfo
  {author} {\bibfnamefont {M.}~\bibnamefont {Sorolla}},\ }\bibfield  {title}
  {\enquote {\bibinfo {title} {Miniaturized coplanar waveguide stop band
  filters based on multiple tuned split ring resonators},}\ }\href {\doibase
  10.1109/LMWC.2003.819964} {\bibfield  {journal} {\bibinfo  {journal} {IEEE
  Microwave and Wireless Components Letters}\ }\textbf {\bibinfo {volume}
  {13}},\ \bibinfo {pages} {511--513} (\bibinfo {year} {2003})}\BibitemShut
  {NoStop}%
\bibitem [{\citenamefont {Fadavi~Roudsari}\ \emph {et~al.}(2023)\citenamefont
  {Fadavi~Roudsari}, \citenamefont {Shiri}, \citenamefont {Renberg~Nilsson},
  \citenamefont {Tancredi}, \citenamefont {Osman}, \citenamefont {Svensson},
  \citenamefont {Kudra}, \citenamefont {Rommel}, \citenamefont {Bylander},
  \citenamefont {Shumeiko},\ and\ \citenamefont {Delsing}}]{Fadavi2023}%
  \BibitemOpen
  \bibfield  {author} {\bibinfo {author} {\bibfnamefont {A.}~\bibnamefont
  {Fadavi~Roudsari}}, \bibinfo {author} {\bibfnamefont {D.}~\bibnamefont
  {Shiri}}, \bibinfo {author} {\bibfnamefont {H.}~\bibnamefont
  {Renberg~Nilsson}}, \bibinfo {author} {\bibfnamefont {G.}~\bibnamefont
  {Tancredi}}, \bibinfo {author} {\bibfnamefont {A.}~\bibnamefont {Osman}},
  \bibinfo {author} {\bibfnamefont {I.-M.}\ \bibnamefont {Svensson}}, \bibinfo
  {author} {\bibfnamefont {M.}~\bibnamefont {Kudra}}, \bibinfo {author}
  {\bibfnamefont {M.}~\bibnamefont {Rommel}}, \bibinfo {author} {\bibfnamefont
  {J.}~\bibnamefont {Bylander}}, \bibinfo {author} {\bibfnamefont
  {V.}~\bibnamefont {Shumeiko}}, \ and\ \bibinfo {author} {\bibfnamefont
  {P.}~\bibnamefont {Delsing}},\ }\bibfield  {title} {\enquote {\bibinfo
  {title} {{Three-wave mixing traveling-wave parametric amplifier with periodic
  variation of the circuit parameters}},}\ }\href {\doibase 10.1063/5.0127690}
  {\bibfield  {journal} {\bibinfo  {journal} {Applied Physics Letters}\
  }\textbf {\bibinfo {volume} {122}},\ \bibinfo {pages} {052601} (\bibinfo
  {year} {2023})}\BibitemShut {NoStop}%
\bibitem [{\citenamefont {Gutierrez-Medina}(2013)}]{Gutierrez2013}%
  \BibitemOpen
  \bibfield  {author} {\bibinfo {author} {\bibfnamefont {B.}~\bibnamefont
  {Gutierrez-Medina}},\ }\bibfield  {title} {\enquote {\bibinfo {title} {Wave
  transmission through periodic, quasiperiodic, and random one-dimensional
  finite lattices},}\ }\href {\doibase 10.1119/1.4765628} {\bibfield  {journal}
  {\bibinfo  {journal} {American Journal of Physics}\ }\textbf {\bibinfo
  {volume} {81}},\ \bibinfo {pages} {104--111} (\bibinfo {year}
  {2013})}\BibitemShut {NoStop}%
\bibitem [{\citenamefont {Collin}(2001)}]{Collin2001}%
  \BibitemOpen
  \bibfield  {author} {\bibinfo {author} {\bibfnamefont {R.~E.}\ \bibnamefont
  {Collin}},\ }\enquote {\bibinfo {title} {Periodic structures and filters},}\
  in\ \href {\doibase 10.1109/9780470544662.ch8} {\emph {\bibinfo {booktitle}
  {Foundations for Microwave Engineering}}}\ (\bibinfo {year} {2001})\ pp.\
  \bibinfo {pages} {550--647}\BibitemShut {NoStop}%
\bibitem [{\citenamefont {Bertuccio}(2022)}]{Bertuccio2022}%
  \BibitemOpen
  \bibfield  {author} {\bibinfo {author} {\bibfnamefont {G.}~\bibnamefont
  {Bertuccio}},\ }\bibfield  {title} {\enquote {\bibinfo {title} {On the
  physical origin of the electro-mechano-acoustical analogy},}\ }\href
  {\doibase 10.1121/10.0009803} {\bibfield  {journal} {\bibinfo  {journal} {The
  Journal of the Acoustical Society of America}\ }\textbf {\bibinfo {volume}
  {151}},\ \bibinfo {pages} {2066--2076} (\bibinfo {year} {2022})}\BibitemShut
  {NoStop}%
\bibitem [{\citenamefont {Dove}(2003)}]{Dove2003}%
  \BibitemOpen
  \bibfield  {author} {\bibinfo {author} {\bibfnamefont {M.~T.}\ \bibnamefont
  {Dove}},\ }in\ \href@noop {} {\emph {\bibinfo {booktitle} {Structure and
  Dynamics: An Atomic View of Materials}}},\ \bibinfo {series and number}
  {\bibinfo {number} {Oxford University Press, Oxford}}\ (\bibinfo {year}
  {2003})\BibitemShut {NoStop}%
\bibitem [{\citenamefont {Maznev}\ and\ \citenamefont
  {Wright}(2014)}]{Maznev2014}%
  \BibitemOpen
  \bibfield  {author} {\bibinfo {author} {\bibfnamefont {A.~A.}\ \bibnamefont
  {Maznev}}\ and\ \bibinfo {author} {\bibfnamefont {O.~B.}\ \bibnamefont
  {Wright}},\ }\bibfield  {title} {\enquote {\bibinfo {title} {Demystifying
  umklapp vs normal scattering in lattice thermal conductivity},}\ }\href
  {\doibase 10.1119/1.4892612} {\bibfield  {journal} {\bibinfo  {journal}
  {American Journal of Physics}\ }\textbf {\bibinfo {volume} {82}},\ \bibinfo
  {pages} {1062--1066} (\bibinfo {year} {2014})}\BibitemShut {NoStop}%
\bibitem [{\citenamefont {Shiri}\ and\ \citenamefont
  {Isacsson}(2019)}]{Shiri2019}%
  \BibitemOpen
  \bibfield  {author} {\bibinfo {author} {\bibfnamefont {D.}~\bibnamefont
  {Shiri}}\ and\ \bibinfo {author} {\bibfnamefont {A.}~\bibnamefont
  {Isacsson}},\ }\bibfield  {title} {\enquote {\bibinfo {title} {An
  electrically controlled heat rectifier using graphene nanoribbons},}\ }in\
  \href {\doibase 10.1109/NMDC47361.2019.9084019} {\emph {\bibinfo {booktitle}
  {2019 IEEE 14th Nanotechnology Materials and Devices Conference (NMDC)}}}\
  (\bibinfo {year} {2019})\ pp.\ \bibinfo {pages} {1--4}\BibitemShut {NoStop}%
\end{thebibliography}%

\end{document}